\documentclass[useAMS,usenatbib]{mn2e}
\usepackage[dvips]{graphicx}
\usepackage{fixltx2e}
\usepackage{epstopdf}
\usepackage{times}
\usepackage{amsmath}
\usepackage{color}
\newcommand{\aap}{A\&A}

\newcommand{\mnras}{MNRAS}

\newcommand{\apj}{ApJ}

\newcommand{\aj}{AJ}

\title[Mass function evolution]{Evolution of the Stellar Mass Function in Multiple-Population Globular Clusters}
\author[E.Vesperini et al.]  {Enrico Vesperini$^1$, Jongsuk Hong$^{2,1}$, Jeremy J. Webb$^{3,1}$, Franca D'Antona$^4$\\ \newauthor and Annibale D'Ercole$^5$\\
  $^1$Department of Astronomy, Indiana University, Bloomington, IN, 47401, USA\\
  $^2$Kavli Institute for Astronomy and Astrophysics, Peking University, Yi He Yuan Lu 5, HaiDian District, Beijing 100871, China\\
  $^3$Department of Astronomy and Astrophysics, University of Toronto, ON, M5S 3H4, Canada\\
$^4$INAF, Osservatorio Astronomico di Roma, I-00040 Monteporzio Catone (Roma), Italy\\
$^5$INAF, Osservatorio Astronomico di Bologna, I-40127 Bologna, Italy\\
}
\begin{document}

\newcommand{\alm}{$\alpha_{0.1-0.5}$}
\newcommand{\aim}{$\alpha_{0.3-0.8}$}
\newcommand{\aum}{$\alpha_{0.5-0.8}$}
\newcommand{\rhratio}{R_{\rm h,1G}/R_{\rm h,2G}}
\newcommand{\msun}{m_{\odot}}

\maketitle

\label{firstpage}

\begin{abstract}
We present the results of a survey of $N$-body simulations aimed at studying the effects of the long-term dynamical evolution on the stellar mass function (MF) of multiple stellar populations in globular clusters.  Our simulations show that if first-(1G) and second-generation (2G) stars have the same initial MF (IMF), the  global MFs of the two populations are affected similarly by dynamical evolution and no significant  differences between the 1G and the 2G MFs arise during the cluster's evolution. If the two populations have different IMFs, dynamical effects do not completely erase memory of the initial differences. Should observations find differences between the global 1G and 2G MF, these  would  reveal the fingerprints of differences in their IMFs.
Irrespective of whether the 1G and 2G populations have the same global IMF or not, dynamical effects can produce differences between the local (measured at various distances from the cluster centre) 1G and 2G MFs; these differences are a manifestation of the process of mass segregation in populations with different initial structural properties.  In dynamically old and spatially mixed clusters, however, differences between the local 1G and 2G MFs can reveal differences between the 1G and 2G global MFs. In general, for clusters with any dynamical age, large differences between the local 1G and 2G MFs are more likely to be associated with differences in the global MF. Our study also reveals a dependence of the spatial mixing rate on the stellar mass,  another dynamical consequence of the multiscale nature of multiple-population clusters.
\end{abstract}


\section{Introduction}
\label{sec:intro}
The discovery of multiple stellar populations in globular clusters has revealed an extremely complex picture of the stellar content of these stellar systems (see e.g. Carretta et al. 2009a, 2009b, Piotto et al. 2015, Milone et al. 2017, Gratton et al. 2012 and references therein). 
Theoretical efforts aimed at answering any of the fundamental questions concerning globular cluster formation, chemical and dynamical evolution must now take into account the implications of the presence of multiple stellar populations and the new challenges  posed by this discovery. 

Examples of some of the issues addressed in the literature include the study of the possible sources of processed gas out of which second-generation (hereafter 2G) stars formed (see e.g. Ventura et al. 2001, Decressin et al. 2007, de Mink et al. 2009, Bastian et al. 2013, D'Ercole et al. 2010, 2012, D'Antona et al. 2016, Denissenkov \& Hartwick 2014), the gas and stellar dynamics during the phases of cluster formation and early evolution (see e.g. D'Ercole et al. 2008, 2016, Bekki 2011, 2017a, 2017b,  Elmegreen 2017) and the possible paths leading to the presence of several distinct 2G populations (see e.g. D'Antona et al. 2016, Bekki et al. 2017), the kinematical differences between different populations (Bellazzini et al. 2012, Richer et al. 2013, Bellini et al. 2015, 2018,  Henault-Brunet et al. 2015, Mastrobuono-Battisti \& Perets 2013, Cordero et al. 2017), differences in the dynamical effects on the binary stars belonging to different populations (D'Orazi et al. 2010, Vesperini et al. 2011, Hong et al. 2015, 2016, Lucatello et al. 2015), the implications of different formation models for the contribution of globular cluster stars to the assembly of the Galactic halo (Vesperini et al. 2010, Martell et al. 2010, 2011, Carretta 2016).

One of the fundamental properties that characterizes a stellar population is its initial stellar mass function (hereafter IMF). The evolution of the stellar mass function (hereafter MF) of globular clusters and its possible dependence on a variety of cluster parameters have been the subject of numerous observational studies (see e.g. Piotto \& Zoccali 1999, De Marchi et al. 2007, Paust et al. 2010, Webb et al. 2017, Sollima \& Baumgardt 2017). On the theoretical side many studies have focussed their attention on  the possible effects of internal dynamical evolution, mass segregation, and star loss on the evolution of the MF and the link between the IMF and the present-day mass function (hereafter PDMF) (see e.g. Vesperini \& Heggie 1997, Baumgardt \& Makino 2003, Trenti et al. 2010, Webb \& Leigh 2015, Webb \& Vesperini 2016 and references therein).

As shown in a number of theoretical studies, mass segregation can produce a significant dependence of the PDMF on the radial distance from a cluster centre and,  during a cluster's dynamical evolution, the preferential escape of low-mass stars flattens the low-mass end of the global MF. Establishing a connection between the PDMF and the IMF and disentagling the possible role of formation and dynamics in  cluster-to-cluster differences in the PDMF and in the radial variations of the PDMF inside individual clusters requires a complete modeling of the dynamical effects on the stellar MF (see e.g. Webb \& Vesperini 2016, Webb et al. 2017).

All the theoretical and observational studies of the globular cluster stellar MF have been carried out in the context of single-population clusters; the only observational study of the MF of multiple populations has been carried out for NGC2808 by Milone et al. (2012a); the data of that investigation cover a narrow stellar mass range spanning different values of stellar masses for the three populations identified at the time in NGC2808. With the advent of JWST the identification of multiple populations down to the bottom of the main sequence  will allow for more comprehensive observational investigations to be carried out leading to a detailed characterization of the MF of different stellar populations.

On the theoretical side the discovery of multiple populations raises new and fundamental questions: do different stellar populations form with same IMFs?  What are the effects of dynamical evolution on the global and the local (measured at different distances from the cluster centre) stellar MF of different stellar populations? 
To address the first question, hydro simulations following in detail the formation of individual stars are needed. In this paper we focus instead on the second question.
All multiple-population globular cluster formation models  agree that 2G stars should initially be more centrally concentrated than 1G stars (see e.g. D'Ercole et al. 2008, Decressin et al. 2007); a number of observational studies have shown that in several Galactic clusters the initial differences in the 2G and 1G spatial distributions predicted by formation models have not been completely erased by dynamical evolution and  2G stars are still more centrally concentrated than 1G stars (see e.g. Bellini et al. 2009, Lardo et al. 2011, Nataf et al. 2011, Carretta et al. 2010, Johnson \& Pilachowski 2012, Milone et al. 2012b, Cordero et al. 2014, Li et al. 2014, Simioni et al. 2016; see also Dalessandro et al. 2014, Nardiello et al. 2015 for some examples of clusters in which the  populations are now completely mixed).

The effects of having a centrally concentrated 2G population on the evolution of a cluster's MF have yet to be considered. If the relative initial properties of the 1G and 2G populations leave an imprint on the PDMFs of globular cluster's than future observational studies will be able to constrain the formation mechanism behind multiple populations.
The specific questions we will address in this paper are the following: assuming that 2G and 1G stars form with the same IMF, do differences in their initial spatial distributions lead to differences in the dynamical effects on their stellar MFs? If we instead assume the 2G and 1G populations do not have the same IMFs, are the initial differences in the MF gradually erased by dynamical evolution or are they, to some extent, preserved in the PDMF?

The outline of the paper is the following. In section 2 we describe our methods and initial conditions. In Section 3 we present our results. In section 4 we conclude with a summary and discussion of our results.
\section{Methods and Initial Conditions}
The results presented in this paper are based on $N$-body simulations run
with the GPU-accelerated version of the code {\sc nbody6} (Aarseth 2003, Nitadori \& Aarseth 2012).
Clusters are assumed to move on circular orbits in the host galaxy tidal field  modeled as a point-mass. For all the models studied the initial ratio of the half-mass radius to the Jacobi radius is equal to about 0.03-0.05.
All our simulations start with 50,000 stars and stars moving beyond a distance from the cluster centre  larger than two times the tidal radius are removed from the simulation.
\begin{figure}
\centering{
  \includegraphics[width=6.5cm]{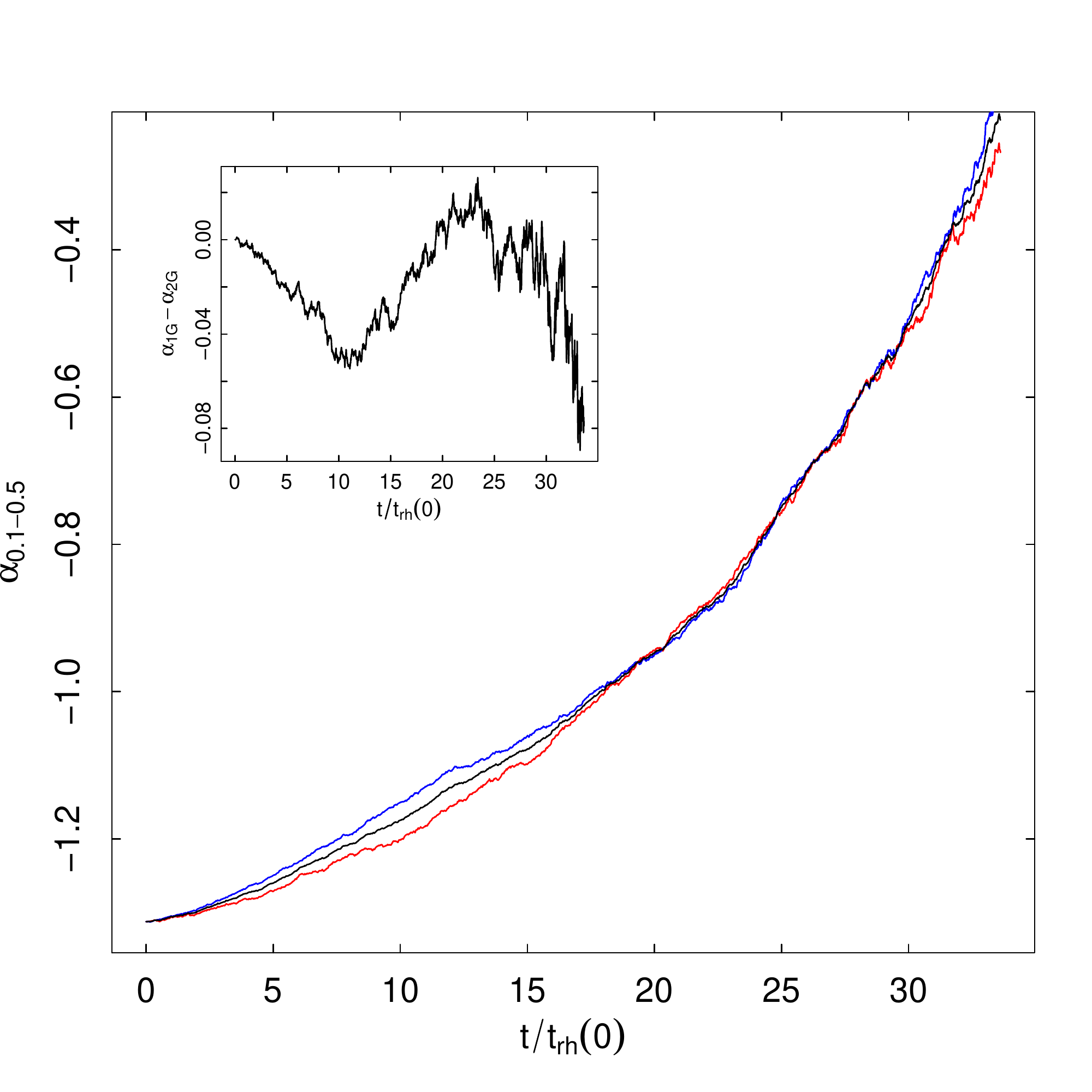}
  \includegraphics[width=6.5cm]{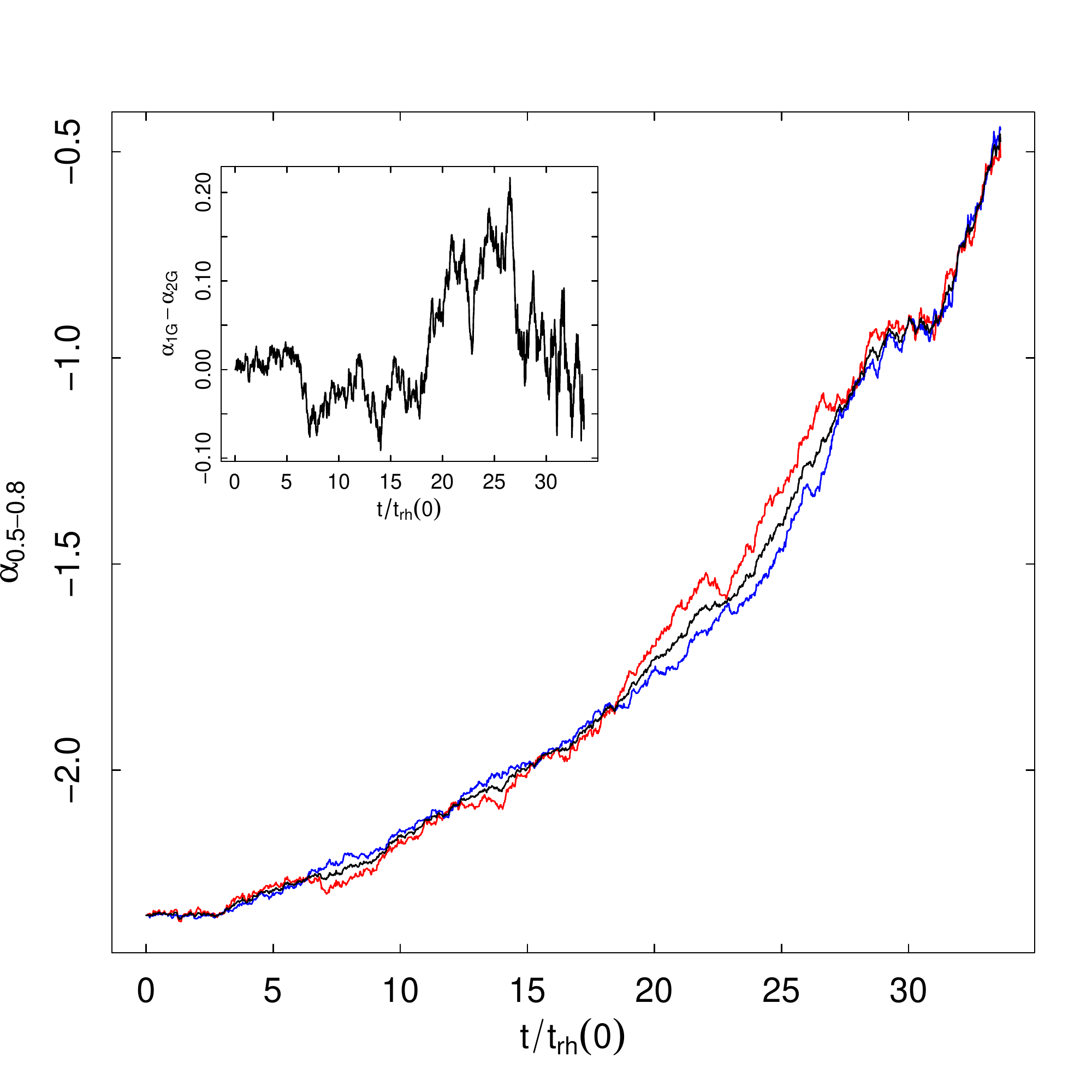}
  \includegraphics[width=6.5cm]{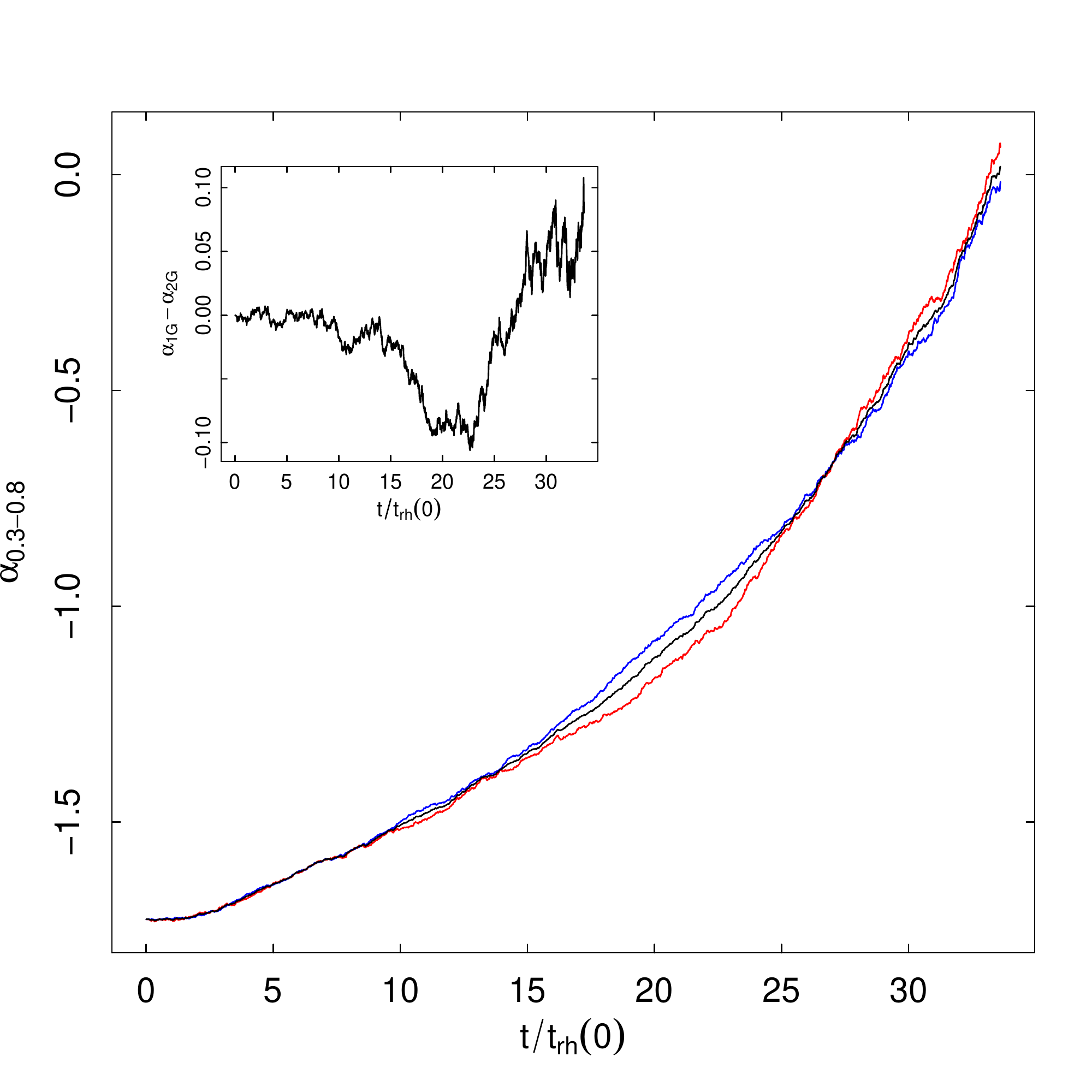}
}
  \caption{Time evolution of the slope of the global MF for  2G stars (blue line), 1G stars (red line) and all stars (black line) for the K01R5 model. The top panel shows the evolution of the slope of the MF for stars with masses between $0.1 \msun$ and $0.5 \msun$; the middle panel shows the evolution of the slope of the MF for stars with masses between $0.5 \msun$ and $0.8 \msun$; the lower panel shows the evolution of the slope of the MF for stars with masses between $0.3 \msun$ and $0.8 \msun$. 
The insets show the time evolution of the difference between the slope of the MF of 1G stars and that of 2G stars for the mass range corresponding to each panel. Time is shown in units of the initial half-mass relaxation time of the entire system.
}
  \label{fig1}
\end{figure}

In all the systems explored, the 1G and 2G subsystems start with the same total number of stars. For each subsytem the density profile is that of a King model (1966) with central dimensionless potential
$W_0=7$ but the 2G subsystem is initially concentrated within the inner regions of the 1G system; we have explored one model in which the initial
ratio of the 3D half-mass radius of the 1G population to that of the 2G population ($\rhratio$) is equal to 5, and one in which this ratio is equal to 10.
 Our initial conditions do not include primordial binaries; a study including primordial binaries will be presented in a future paper.

To study the effects of initially identical IMFs, 1G and 2G star masses are both distributed according to a Kroupa (2001) IMF between 0.1 $m_{\odot}$ and 100 $m_{\odot}$ initially evolved to 11.5 Gyr using the McLuster software (Kuepper et al. 2011) with the stellar evolution models of Hurley et al. (2000) for a metallicity $Z=10^{-3}$ . Neutron stars and black holes are not retained in the cluster after this initial stellar evolution step. 
We refer to the models with a Kroupa (2001) IMF and $R_{\rm h,1G}/R_{\rm h,2G}=5$ and $R_{\rm h,1G}/R_{\rm h,2G}=10$, respectively, as K01R5 and K01R10.

In addition we have explored the evolution of three systems in which the IMF 1G stars with masses between 0.1 $m_{\odot}$ and 0.5 $m_{\odot}$ has a power-law index, $\alpha$, equal to 1.0, 0.8, and 0.5 (for a Kroupa (2001) IMF $\alpha$ for stars between 0.1 and 0.5 $m_{\odot}$ is equal to 1.3) while 2G stars have a standard Kroupa (2001) IMF. We refer to these simulation as K01R5a1, K01R5a08, and K01R5a05. Finally we have also studied the evolution of  a cluster in which 2G stars with masses between 0.1 $m_{\odot}$ and 0.5 $m_{\odot}$ are distributed according to a IMF with a power-law index equal to 0.5 while 1G stars have a Kroupa (2001) IMF; we refer to this simulation as K01R5-2ga05.
  For these additional simulations we have adopted a ratio of the 1G to the 2G half-mass radius ($R_{\rm h,1G}/R_{\rm h,2G}$) equal to 5. 

In order to characterize the evolution of the stellar MF, we fit the MF of main sequence stars between $m_{min}$ and $m_{max}$ with a power-law function $dN/dm \propto m^{-\alpha}$ and use a maximum-likelihood calculation (see e.g. Trenti et al. 2010) to calculate $\alpha$. We report the time evolution of $\alpha$ for a few different pairs of values for $(m_{min}, m_{\max})$: $(0.1,0.5)\msun$, $(0.3,0.8)\msun$, $(0.5,0.8)\msun$. Hereafter we refer to the power-law index $\alpha$ as the slope of the MF and we indicate the values of  $\alpha$ for the three mass ranges explored, respectively,  as $\alpha_{0.1-0.5}$, $\alpha_{0.3-0.8}$, $\alpha_{0.5-0.8}$.
All the simulations were run until about 80 per cent of the initial mass was lost.

\section{Results}
\subsection{Evolution of the global mass function}
\begin{figure}
\centering{
  \includegraphics[width=6.5cm]{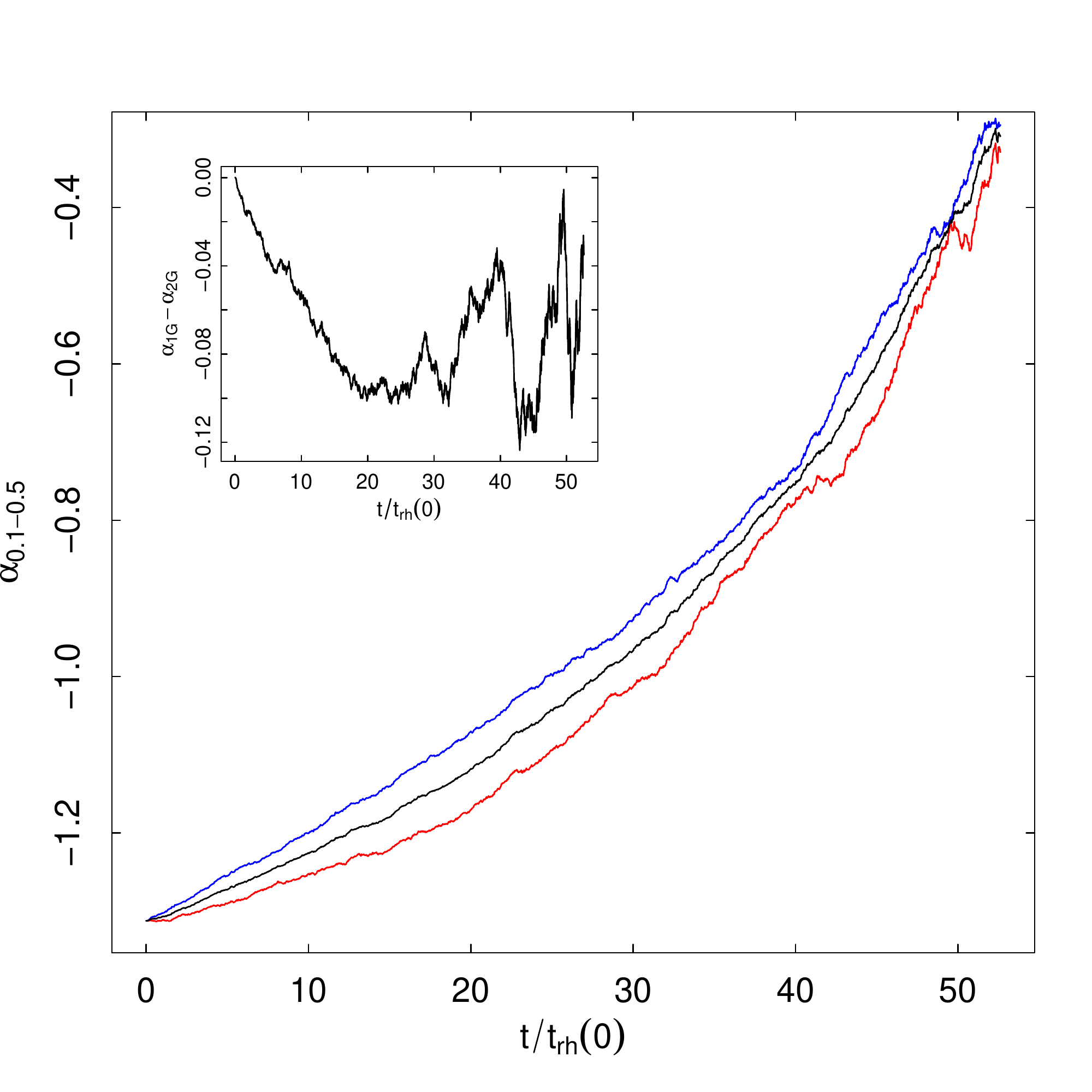}
  \includegraphics[width=6.5cm]{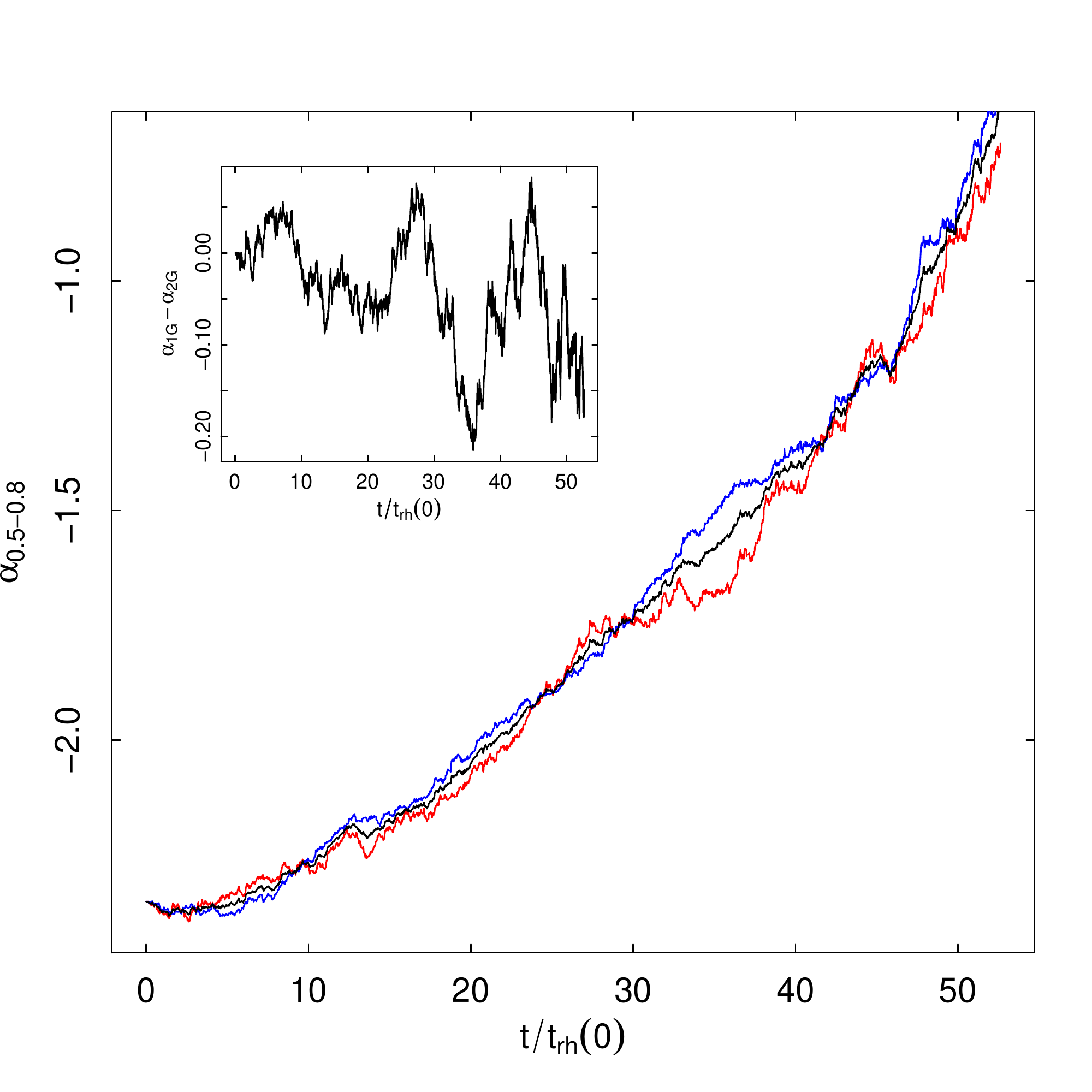}
  \includegraphics[width=6.5cm]{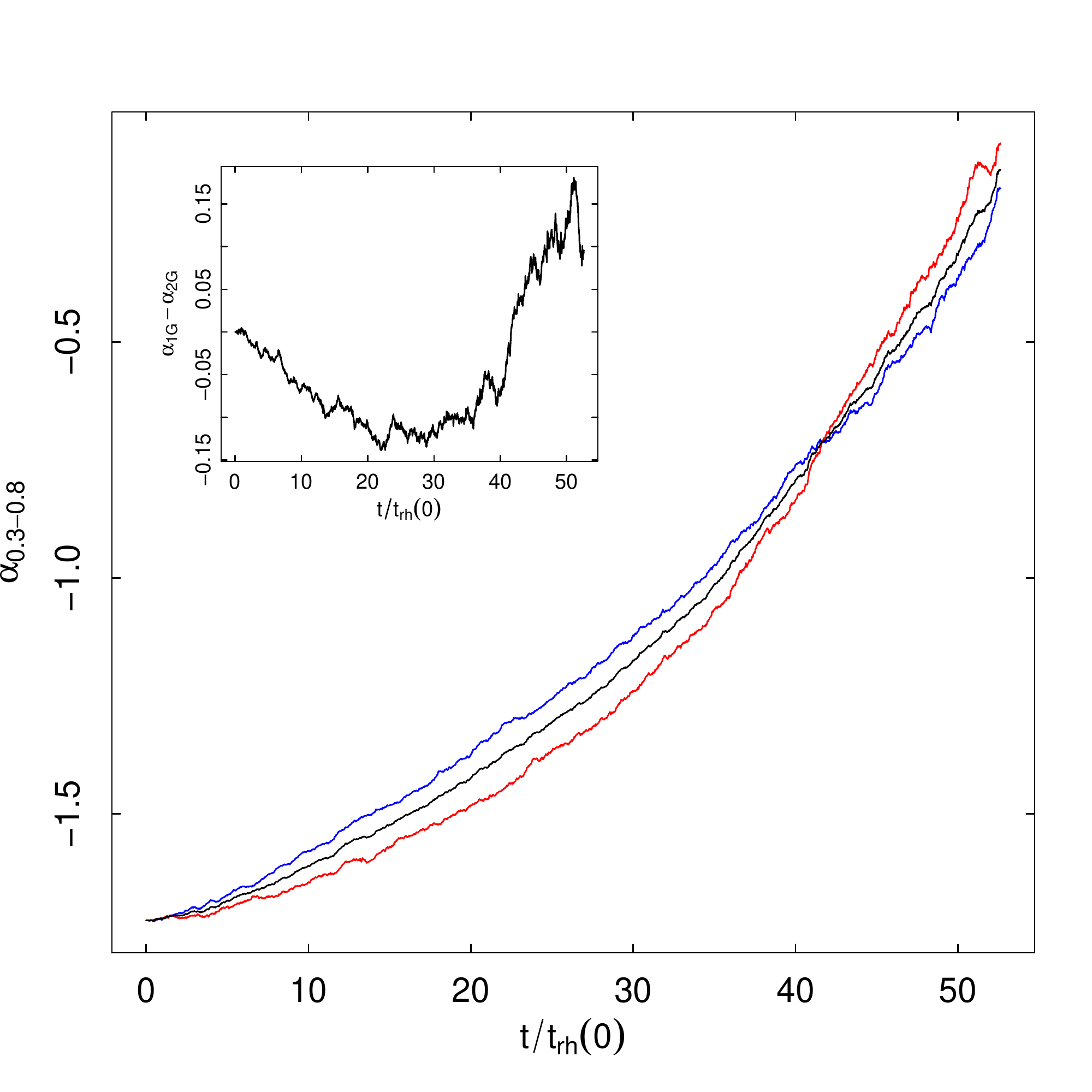}
}
  \caption{Same as Figure 1 but for the model K01R10
}
  \label{fig2}
\end{figure}

We start the presentation of our results by comparing the evolution of the slope of the global 1G and 2G MF for the K01R5 and K01R10 models. In Fig. \ref{fig1} we show the time evolution (with time normalized to the initial half-mass relaxation time of the entire system) of \alm, \aim, and \aum$~$ for the 1G and the 2G populations for the simulation K01R5. In all cases, the evolution of the slopes of the global MFs of the 1G and the 2G is very similar and only negligible differences between the MF of the 1G and the 2G populations arise during the cluster evolution. In Fig. \ref{fig2} we plot the time evolution of  \alm, \aim, \aum$~$, for the model K01R10 and show that this conclusion does not depend on the initial concentration of the 2G population. 

\begin{figure}
\centering{
  \includegraphics[width=6.8cm]{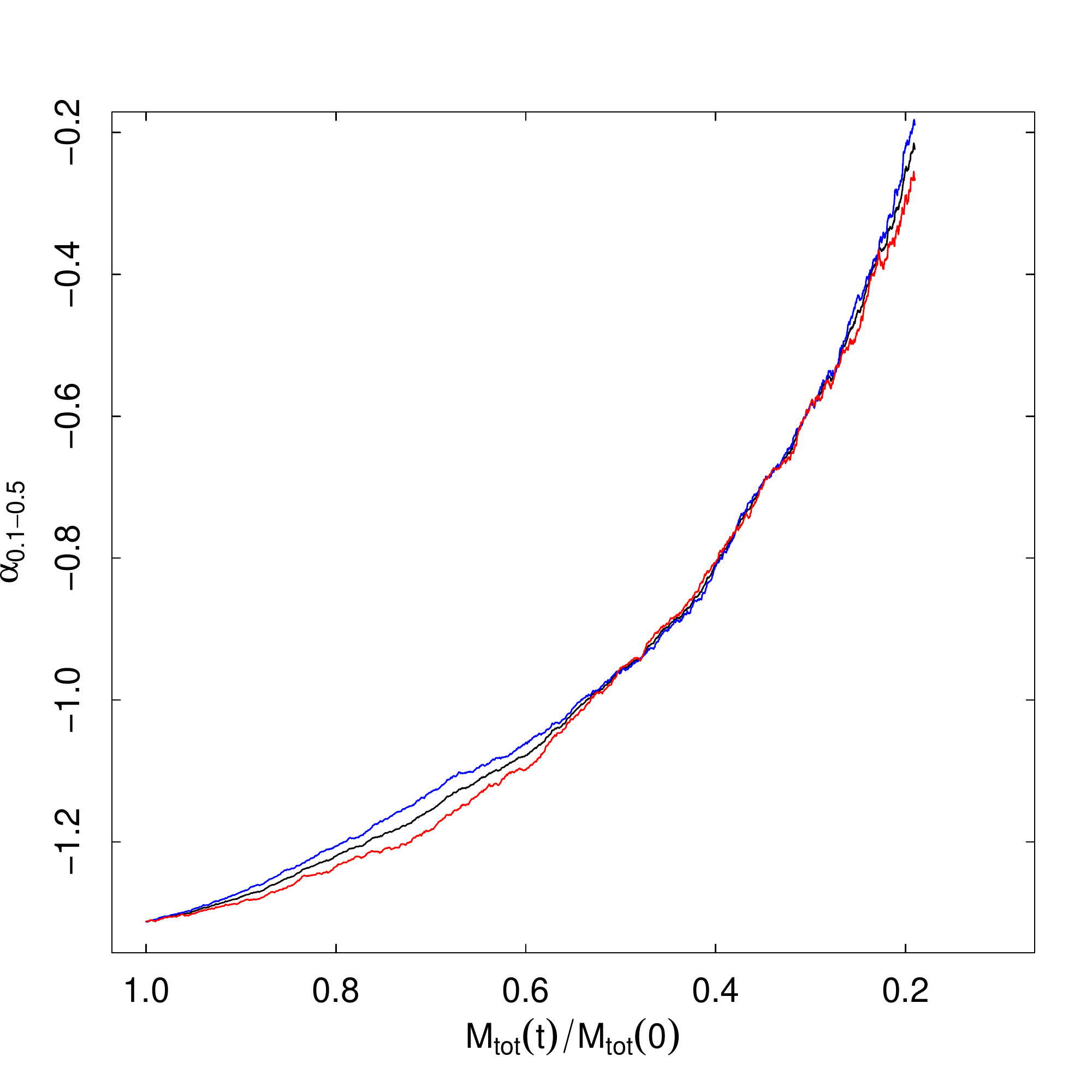}
  \includegraphics[width=6.8cm]{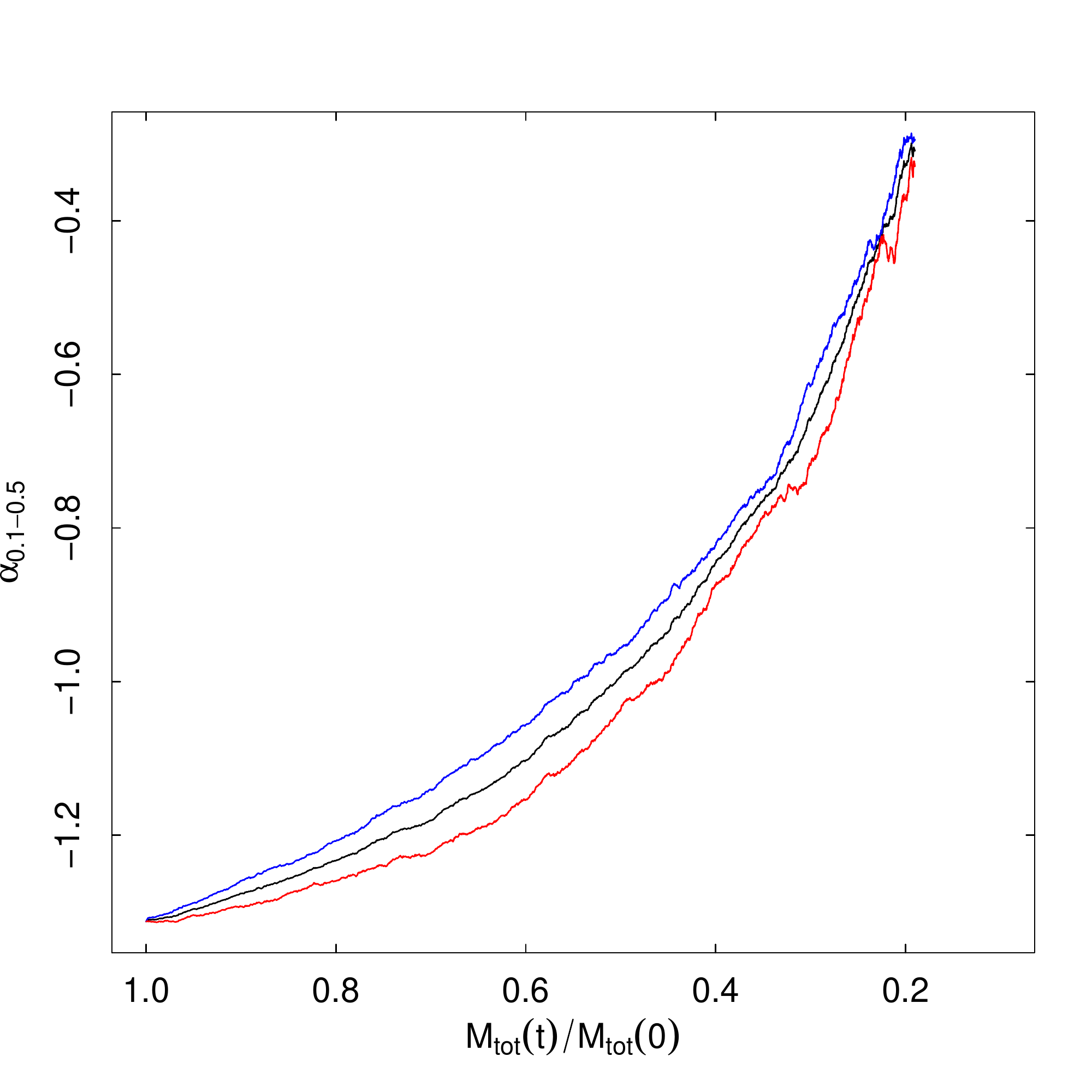}
}
  \caption{Evolution of \alm$~$ for 2G stars (blue line), 1G stars (red line) and all stars (black line) as a function of the fraction of the total initial mass remaining in the cluster for the K01R5 model (top panel) and the K01R10 model (bottom panel).
}
  \label{fig3}
\end{figure}

To understand the small differences in the evolution of the 1G and 2G MFs, it is important to first consider how the MF of a single population cluster evolves.
A number of studies in the literature (see e.g. Vesperini \& Heggie 1997, Baumgardt \& Makino 2003, Trenti et al. 2010) have shown that the evolution of the slope of the global MF is correlated with the amount of star loss due to two-body relaxation (notice, however, that a cluster early expansion triggered, for example, by mass loss due to stellar evolution or primordial gas expulsion can result in a significant loss of stars without affecting the slope of the cluster MF; such early episode of star loss may leave no fingerprint in the slope of the MF).
\begin{figure}
\centering{
  \includegraphics[width=7.5cm]{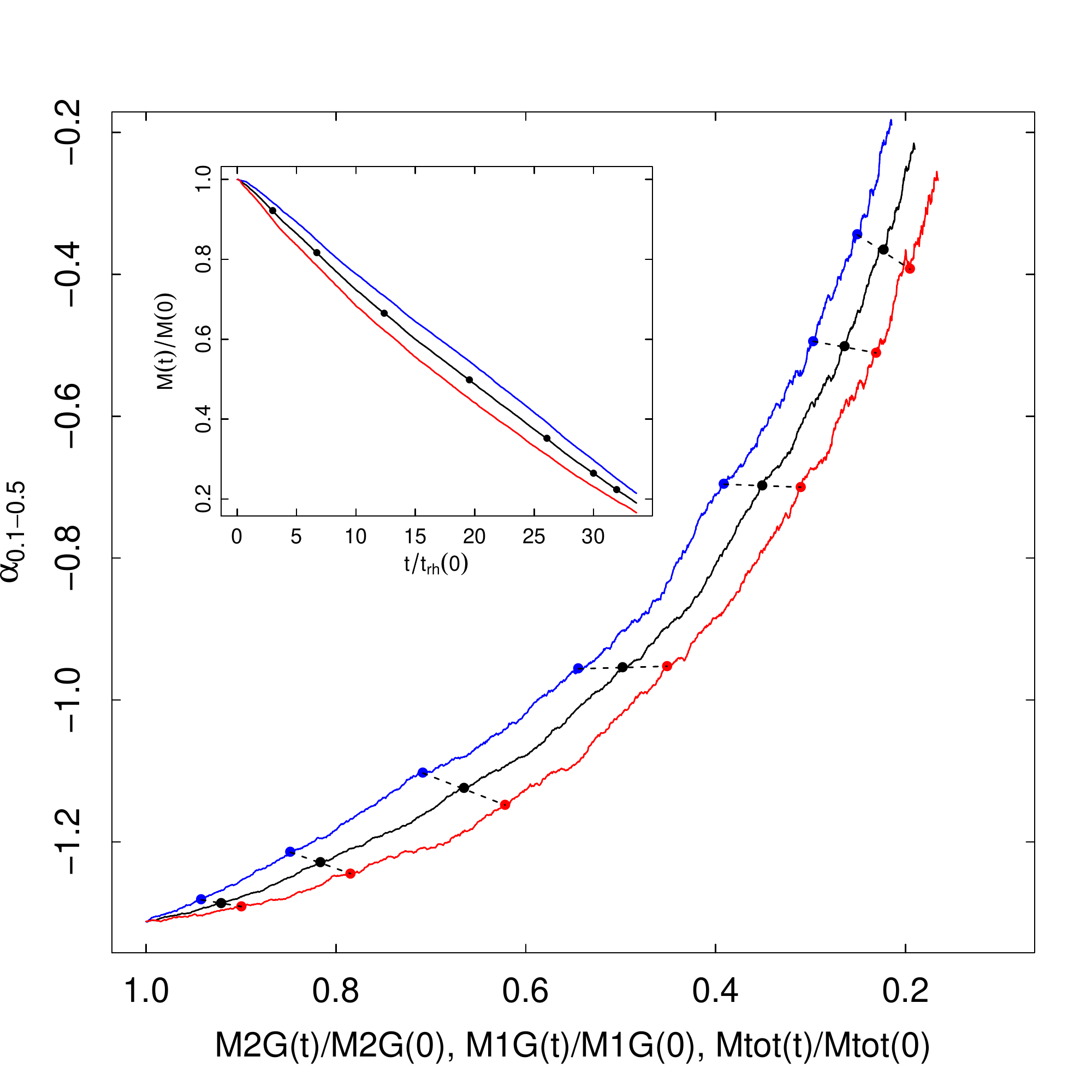}
}
  \caption{Evolution of \alm$~$ for 2G stars (blue line), 1G stars (red line) and all stars (black line) as a function, respectively, of the fraction of the total initial mass remaining in 2G stars, 1G stars an all stars for the K01R5 model. The inset shows the time evolution of the total mass in 2G stars (blue line), 1G stars (red line) and all stars (black line) (each normalized to its initial value). Time is expressed in units of the initial half-mass relaxation time of the entire cluster. In the main panel, dashed segments join the three lines at the few selected times shown as black filled dots in the inset.
}
  \label{fig4}
\end{figure}

In Fig.\ref{fig3} we plot \alm$~$ as a function of the fraction of the initial cluster mass remaining in the cluster, $M(t)/M(0)$, and show that this correlation holds also for multiple-population clusters.
On the other hand, as shown in Fig. \ref{fig4}, the relationship between the slope of the MF and the mass loss of each population is different for the 1G and the 2G  populations. The 2G subsystem is initially more compact and tidally underfilling than the 1G system; this implies that the 2G system can develop a larger degree of mass segregation before starting to lose stars and that a given amount of star loss will lead to a stronger flattening of the MF than that occurring for a less segregated system losing the same fraction of mass (see e.g. Trenti et al. 2010 for differences in the evolution of $\alpha$ versus $M(t)/M(0)$ for filling and underfilling single-population clusters). The small differences between the evolution of $\alpha$ for 1G and 2G stars can be understood in terms of the slight differences in the star loss rate of the two populations. The two populations do not lose stars at the same rate; as already shown in previous studies (see e.g. D'Ercole et al. 2008, Decressin et al. 2008, Vesperini et al. 2013), also during the cluster long-term evolution there is a preferential loss of 1G stars. The slower rate at which the 2G system loses stars is in part compensated by the stronger dependence of $\alpha$ on $M(t)/M(0)$. The interplay beween the different star loss rates and the evolution of $\alpha$ for the two populations is illustrated in Fig.\ref{fig4}: on the lines for the 1G and the 2G populations we have also plotted a few points corresponding to the same value of time. At any given time, the cluster has lost a larger number of 1G stars than 2G stars but the more rapid evolution of $\alpha$ for the 2G population compensates this difference and results in a similar evolution for the MF of the two populations.
\begin{figure}
\centering{
  \includegraphics[width=7.5cm]{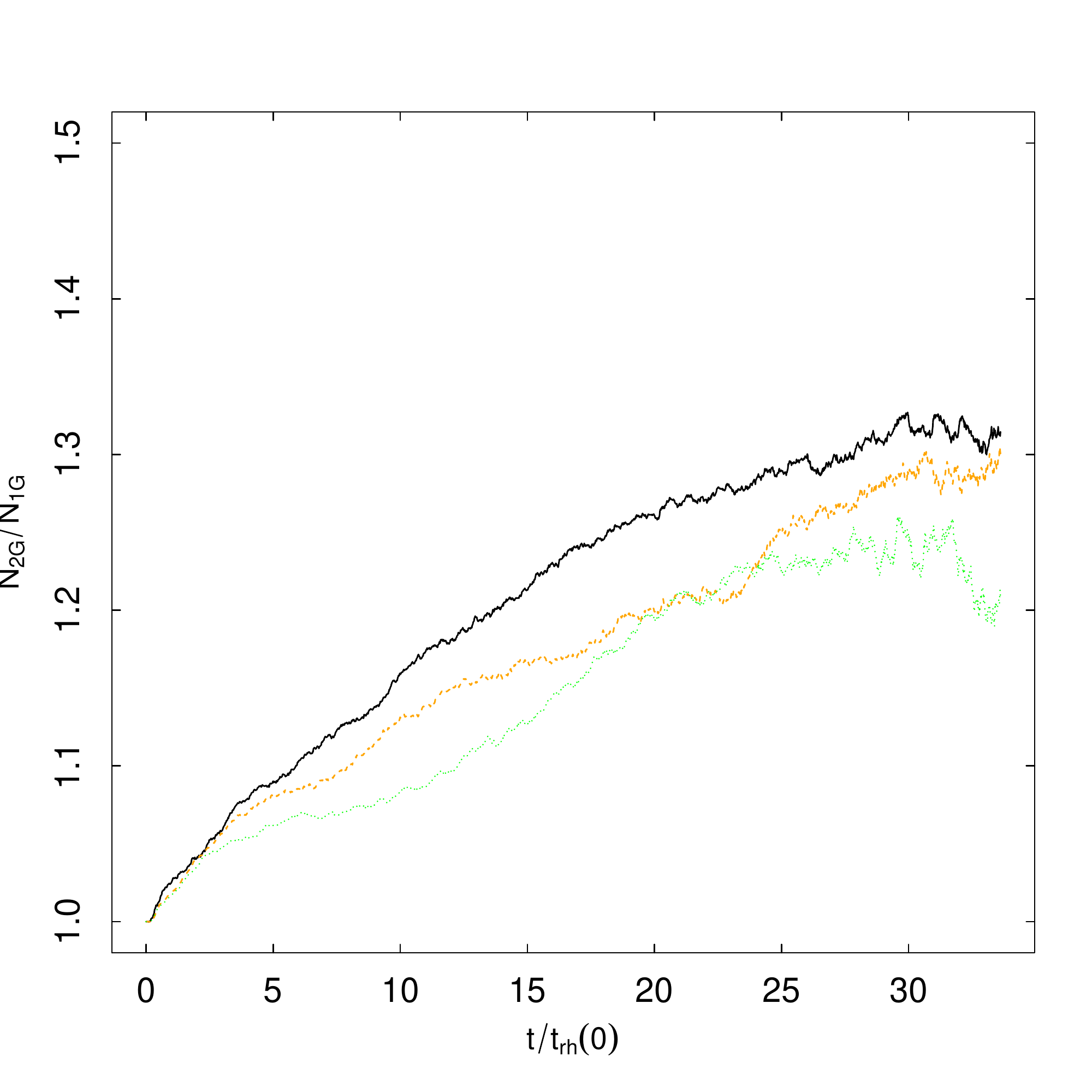}
  \includegraphics[width=7.5cm]{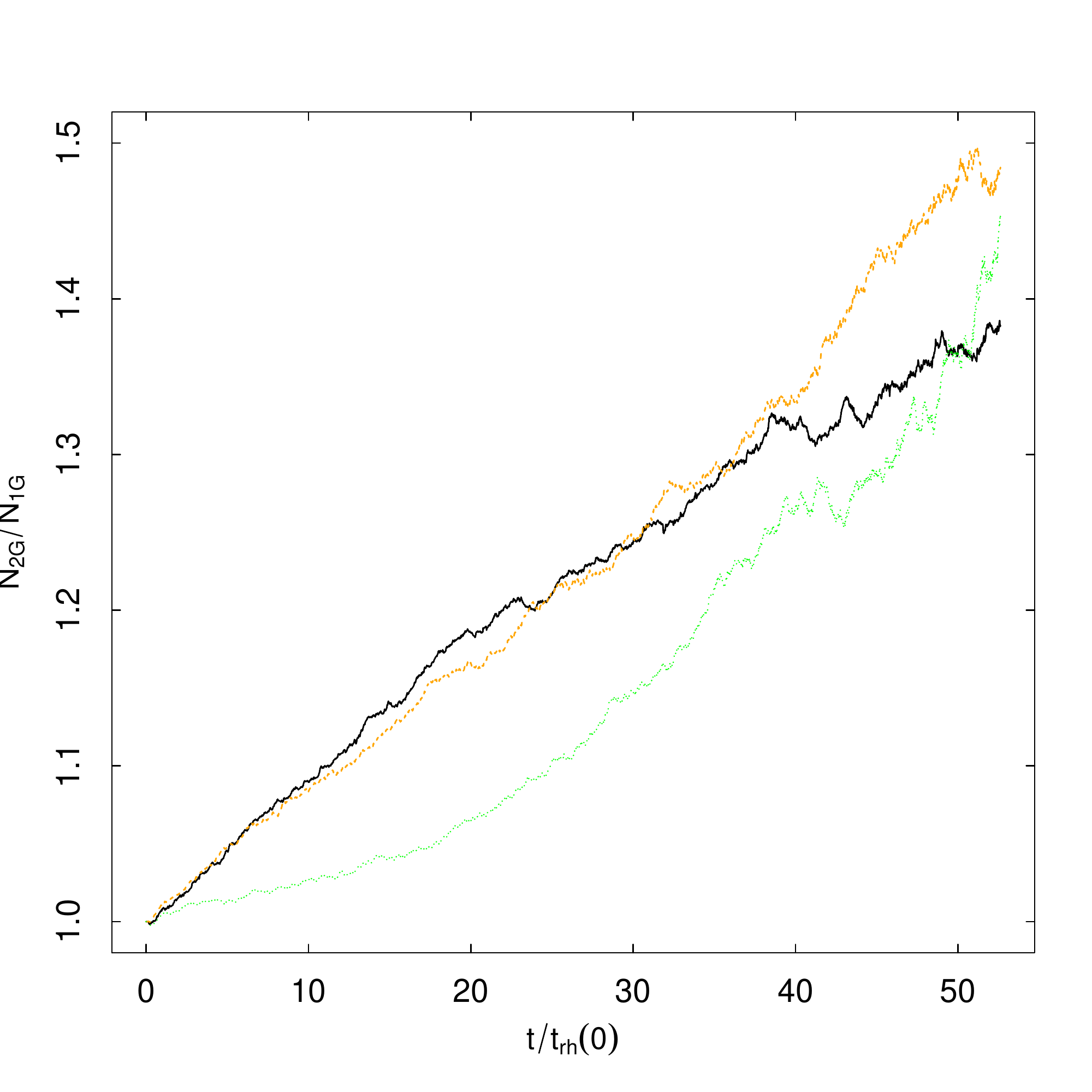}
}
  \caption{Time evolution of the ratio of the total number of 2G stars to the total number of 1G stars with masses in the range ($0.1\msun$-$0.3\msun$; green line), ($0.3\msun$-$0.6\msun$; orange line), and ($0.6\msun$-$0.85\msun$; black lines) for the K01R5 model (top panel) and the K01R10 model (bottom panel). Time is expressed in units of the initial half-mass relaxation time of the entire cluster. 
}
  \label{fig5}
\end{figure}

This point is further illustrated in Fig. \ref{fig5} in which we show the time evolution of the 2G-to-1G number ratio for all stars and for stars in different ranges of mass. Fig \ref{fig5} implies that if a globular cluster starts its long-term evolution with a 2G-to-1G number ratio independent of stellar mass and the same IMF for 1G and 2G stars, star escape due to two-body relaxation will produce only small differences in the 2G-to-1G number ratio  for different stellar mass ranges. 

\subsection{Evolution of the local mass function}
In the previous subsection we have shown that differences in the spatial distribution of 1G and 2G stars do not lead to significant differences in the evolution of the two populations' global MF. A direct observational measure of the global MF, however, is challenging and observational studies often can only determine the local MF within some specific range of distances from the cluster centre. As a cluster evolves, the effects of two-body relaxation lead to the segregation of the more massive stars toward  the cluster's inner regions while the outer regions become increasingly dominated by low-mass stars: this implies that the local slope of the MF is, in general, different from the global one. Here we explore the implications of the differences in the spatial distributions of 1G and 2G stars for the evolution of their local MFs.
\begin{figure*}
\centering{
  \includegraphics[width=7cm]{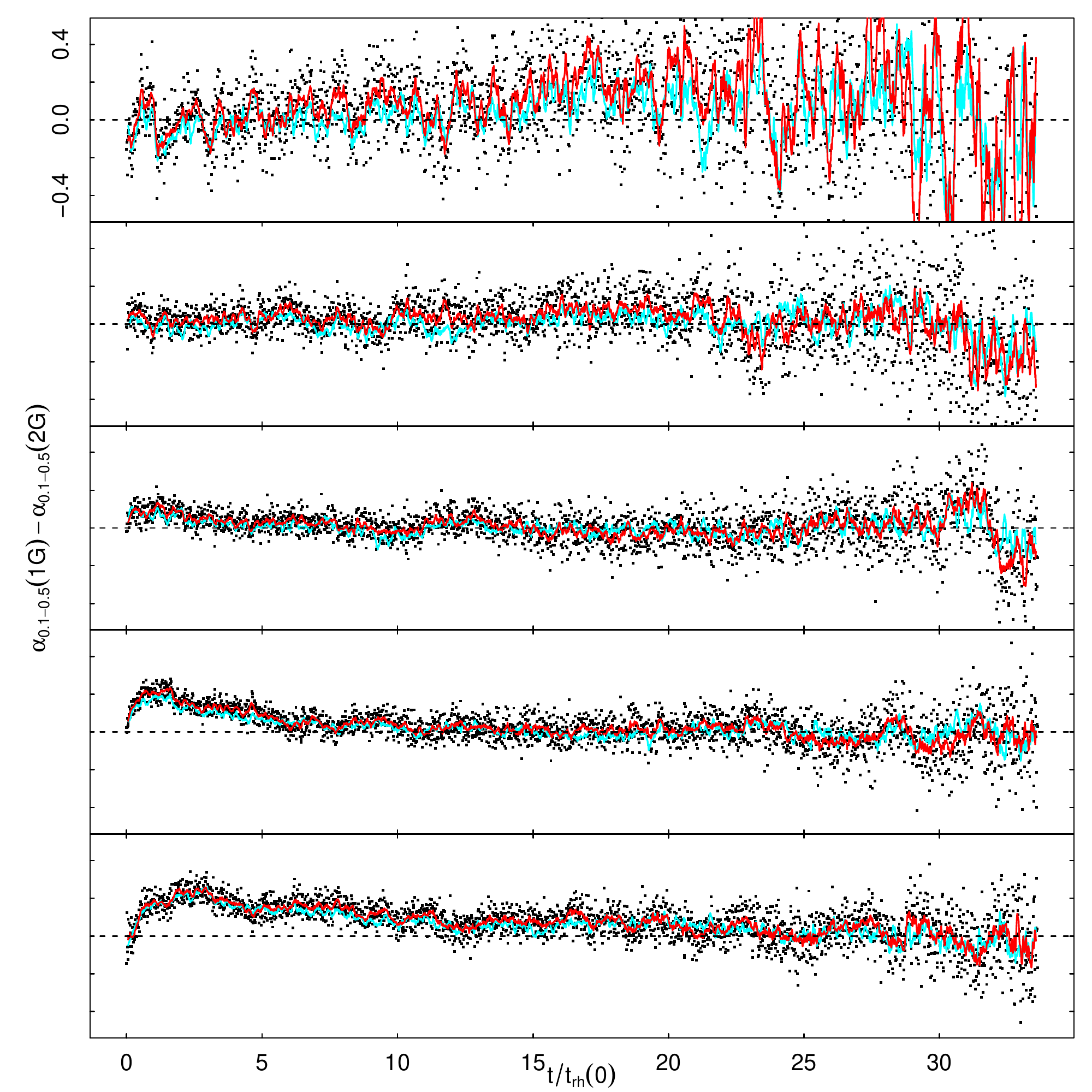}
  \includegraphics[width=7cm]{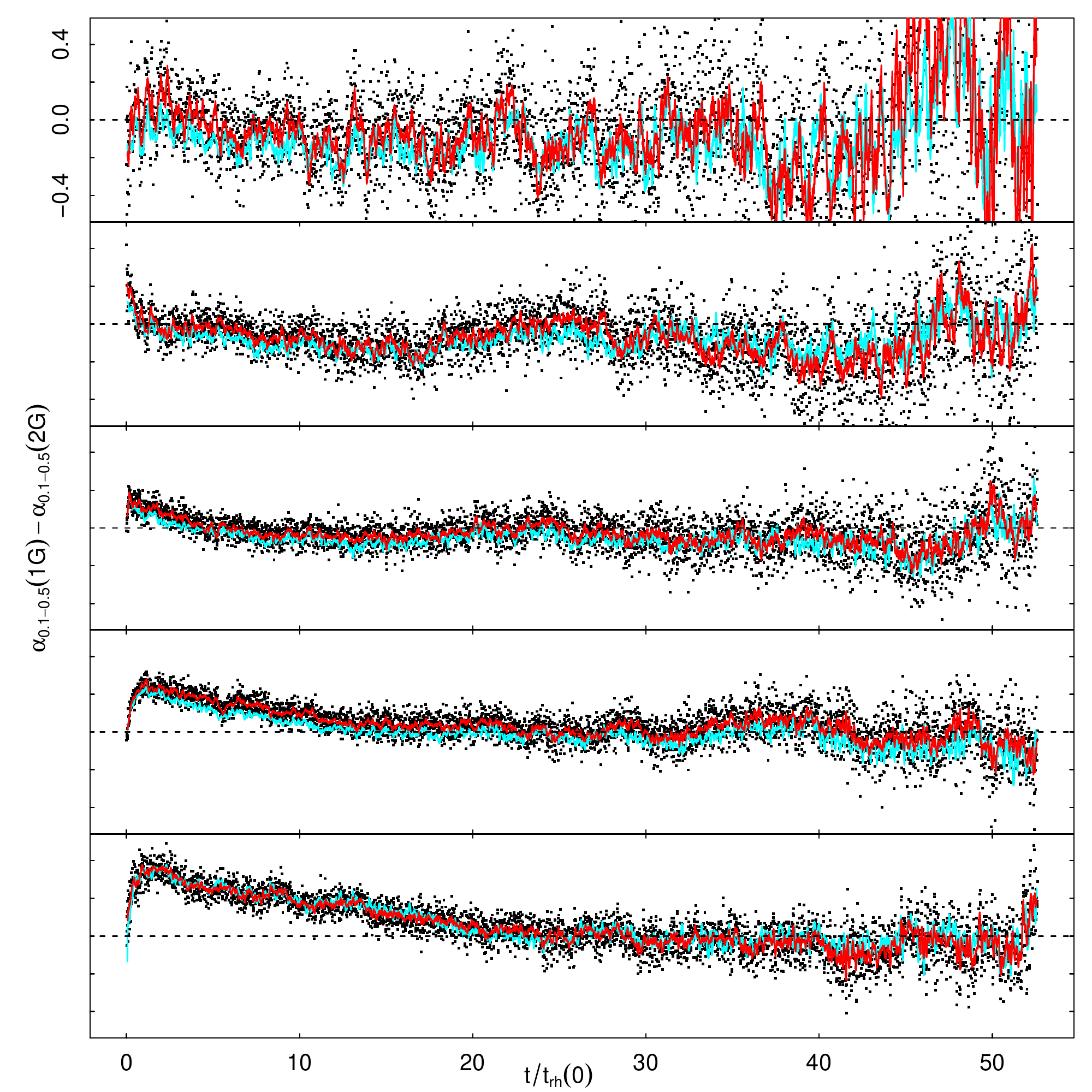}
}\\
\centering{
  \includegraphics[width=7cm]{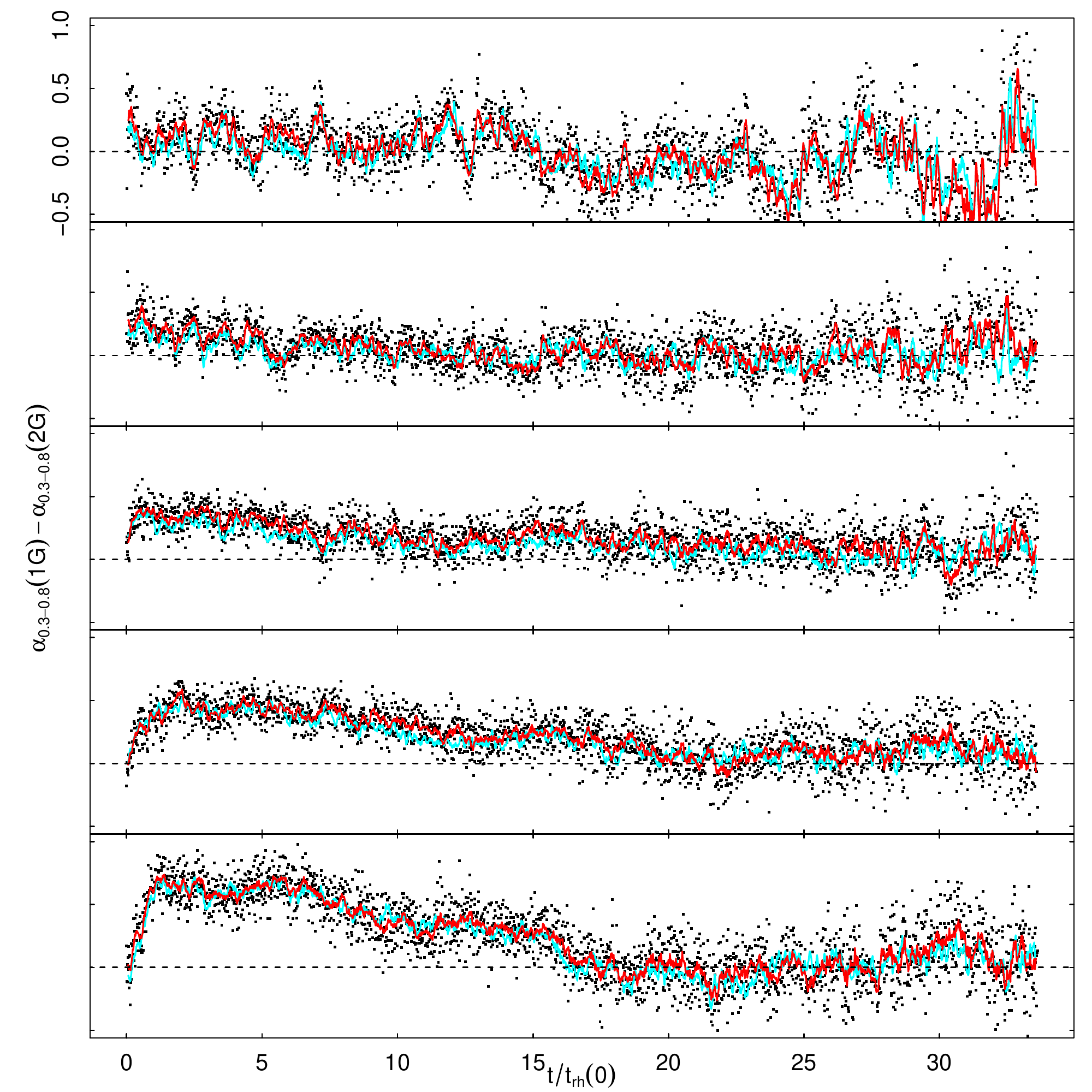}
  \includegraphics[width=7cm]{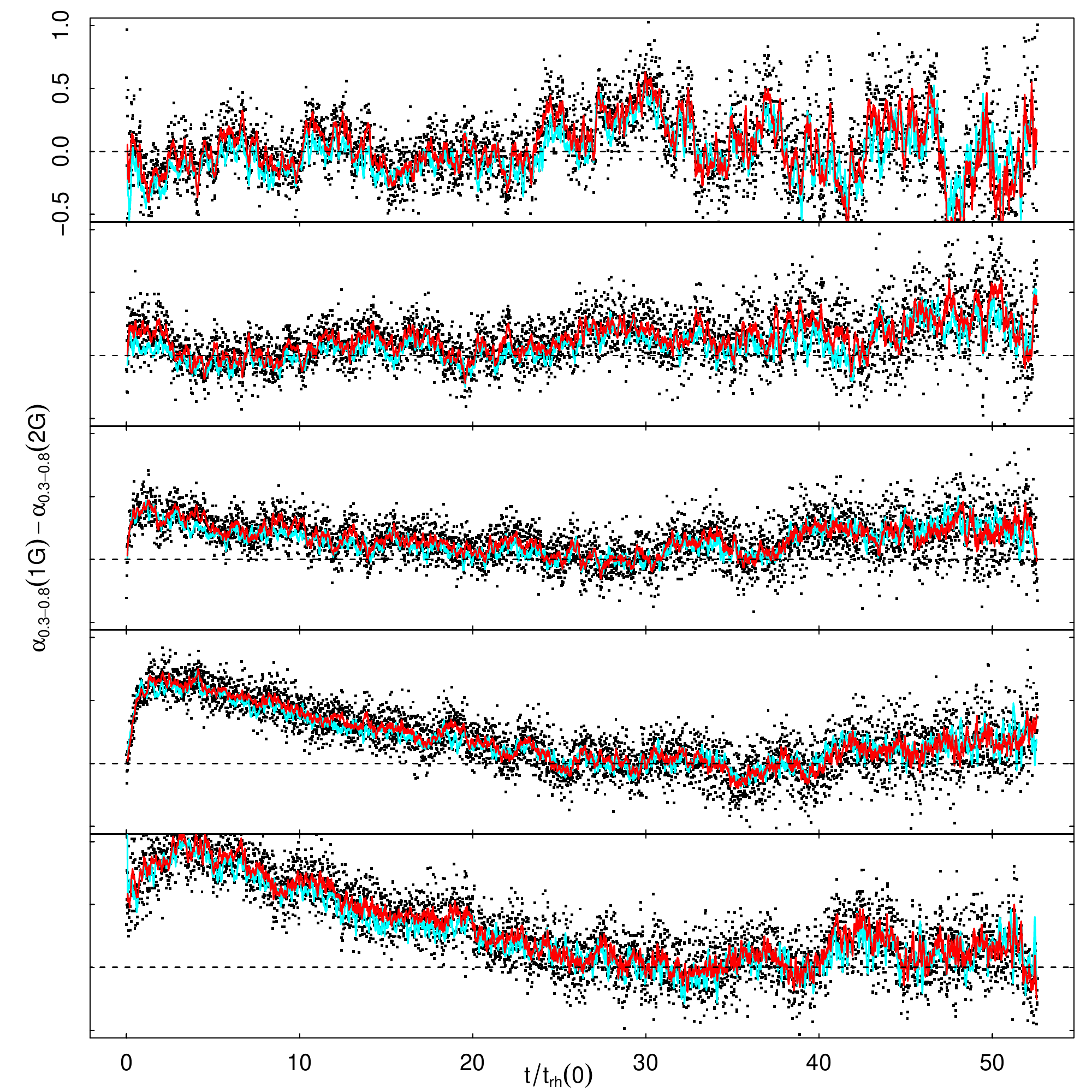}
}\\
\centering{
  \includegraphics[width=7cm]{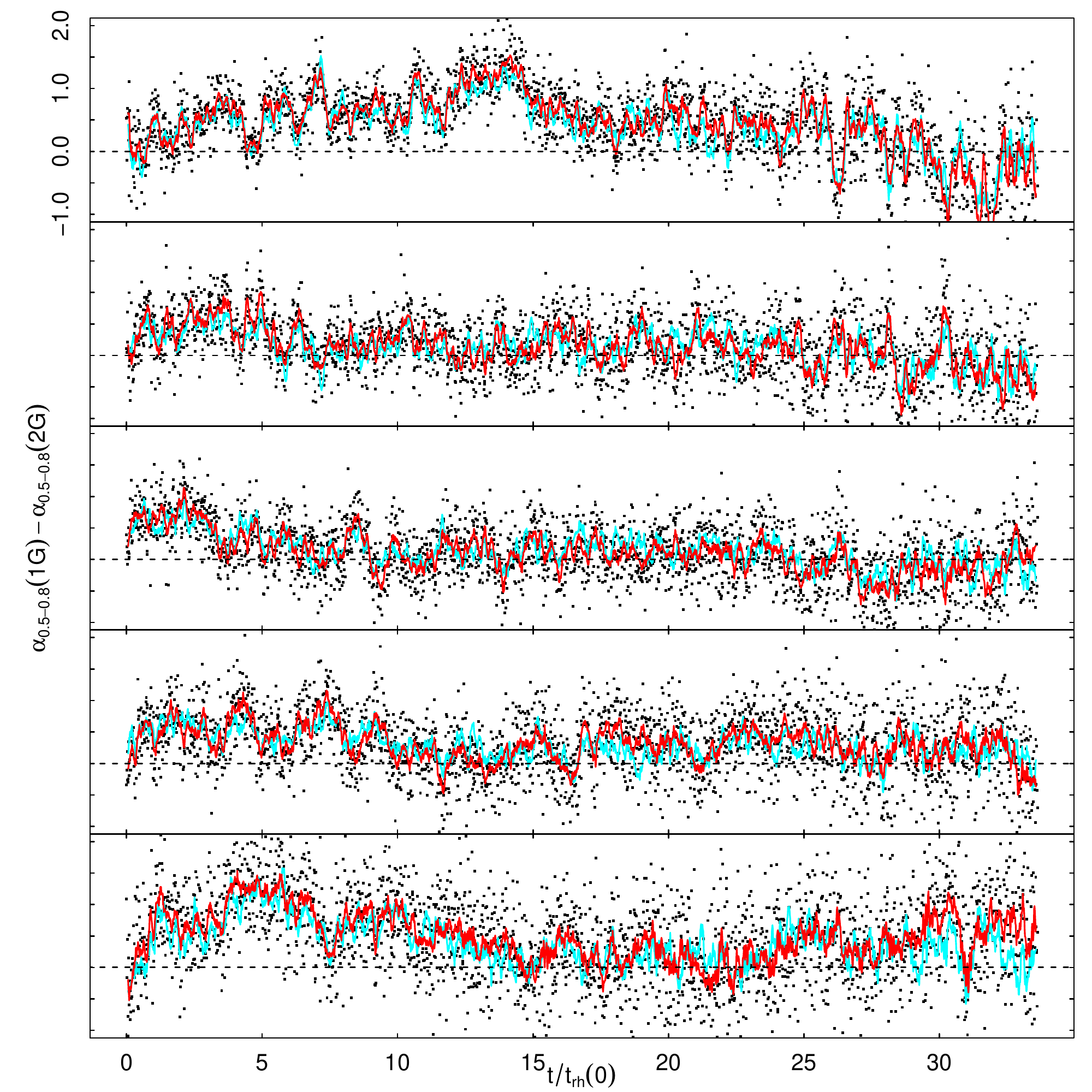}
  \includegraphics[width=7cm]{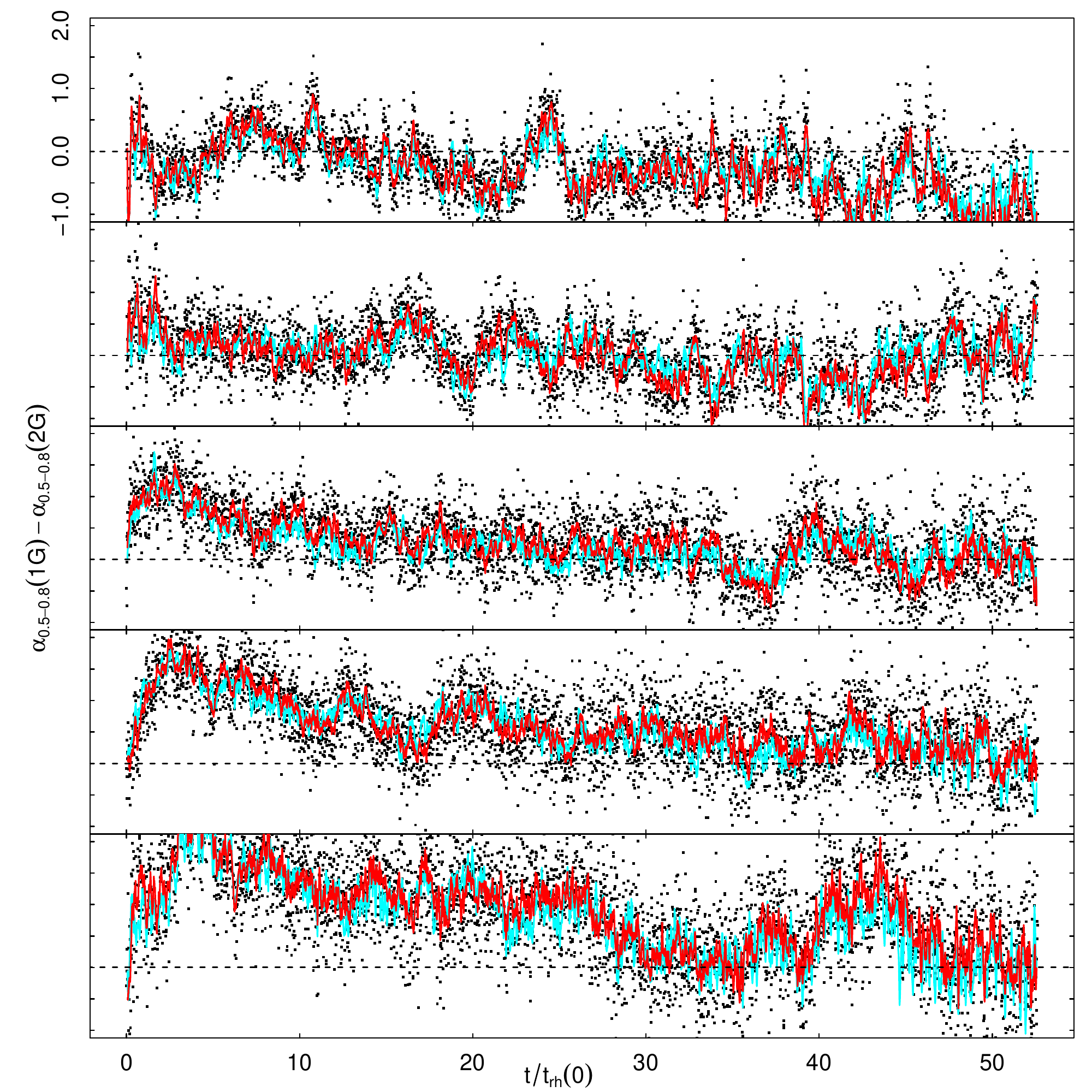}
}\\
  \caption{Time evolution of the difference between the 1G and the 2G slope of the mass function in the mass range ($0.1$-$0.5$)$\msun$ (top row), ($0.3$-$0.8$)$\msun$ (middle row), ($0.5$-$0.8$)$\msun$ (bottom row) in shells at different distances from the cluster centre for the K01R5 model (lefth-hand panels) and the K01R10 (right-hand panels). The five panels in each figure correspond, from the top to the bottom panel of each figure, to 3D (black points and red lines) and 2D (cyan lines) radial shells limited by the following lagrangian radii (0\%-10\%), (10\%-25\%), (25\%-45\%), (45\%-65\%), (75\%-90\%).
 Red lines show the rolling means (calculated using 10 points) of the difference between the 1G and 2G  MF slopes calculated in 3D lagrangian radii shells and $t/t_{\rm rh}(0)$; cyan lines show the rolling means (calculated using 10 points) of the difference between the 1G and 2G  MF slopes calculated in 2D lagrangian radii shells  and $t/t_{\rm rh}(0)$.
 Time is expressed in units of the initial half-mass relaxation time, $t_{\rm rh}(0)$ of the entire cluster. 
}
  \label{fig6}
\end{figure*}

In Fig. \ref{fig6} we show the time evolution of the difference between the 1G and 2G values of \alm, \aim, and \aum$~$ measured at various 3D and 2D distances from the cluster centre. While, as discussed in the previous subsection, the global MFs of the two populations remain very similar to each other during the entire cluster's  evolution, the locally measured slopes of the MFs may be characterized by significant differences. The initial differences in the 1G and 2G spatial distributions imply that the relaxation timescale for the 2G subsystem is shorter than that of the 1G system and the process of segregation of more massive stars towards the inner regions and diffusion of lighter stars towards the outer regions proceeds more rapidly for 2G stars. The evolution of the differences between the 1G and 2G local MFs is therefore driven by the differences in the initial structural properties of the two populations and closely connected with the evolution towards complete mixing of their spatial distributions. The initial differences in the spatial distributions of the two populations imply that a given region of the cluster might be populated by 1G and 2G stars in different ways: for example, the intermediate and outer cluster's shells are initially populated mainly by 1G stars and only later become increasingly populated with 2G stars as they gradually diffuse towards the outer regions and mix with 1G stars. The 2G stars diffusing in the outer regions and gradually mixing with the 1G population formed there are initially dominated by low-mass stars as two-body relaxation drives massive stars towards the inner regions while low-mass stars migrate outwards: this implies that the local MF of 2G stars in the cluster's outer regions is steeper than that of the 1G and explains the initial increase in the difference between the 1G and 2G local slope of the MF in the intermediate and outer shells shown in Fig. \ref{fig6} (see the panels corresponding to the shells limited by the (45\%-65\%) and (75\%-90\%) lagrangian radii).

The plots in Fig. \ref{fig6} illustrate that care will be needed in the interpretation of future observations of the local MF of multiple populations: observed differences in the local slope of the MF may be associated with differences in the global MF of  1G and 2G stars (see section \ref{sec:imf} below) or, as is the case in the simulations presented here, to the dynamics of  populations with different initial spatial distributions. As we will further discuss in the next section, the extent of the observed difference between the local 1G and 2G MF may provide an indication on whether they are due only to dynamical effects or also to differences between the global MFs.
Fig.\ref{fig6} illustrates this point for some specific mass ranges and lagrangian shells. In order to make a stronger connection to observational studies, the extent of the differences in the MF explored in future observations will require  an analysis focussing on the specific  mass range and distances from the cluster centre covered by the observations.

 \subsection{Stellar populations with different IMFs}
\label{sec:imf}
In the previous sections we have assumed that the 1G and 2G populations form with the same IMF. While this may be a reasonable assumption, it is possible that the two populations might actually be characterized by different IMFs. 
In this section we explore the evolution of the MF of clusters in which 1G and 2G stars do not have the same IMF.  Specifically, we study three systems with a 1G slope for stars with masses between $0.1 \msun$ and $0.5\msun$ initially equal to \alm$=1.0$, \alm$=0.8$, and \alm$=0.5$ (for the 2G population we keep the standard slope of the Kroupa 2001 IMF \alm$=1.3$) and one system in which the 1G population has a Kroupa (2001) IMF and the 2G's IMF slope for stars with masses between $0.1 \msun$ and $0.5\msun$ is equal to 0.5.

We emphasize that, at this stage, neither observational nor theoretical star formation studies provide any guidance about whether the IMF of the two populations might actual differ and the possible extent of these differences; a complete study of this issue without any indication of plausible choices for the 1G and 2G IMFs would therefore require a much more extensive survey of simulations exploring a broad range of possible choices for the IMFs of the two populations. Here we provide a much more limited initial study: the goal of the four simulations presented  is to carry out an initial exploration of 1) how differences in the MFs of two populations might evolve as a result of the effects of a cluster dynamical evolution and 2) of the extent to which memory of the initial differences is preserved during a cluster's evolution.
\begin{figure*}
\centering{
  \includegraphics[width=5cm]{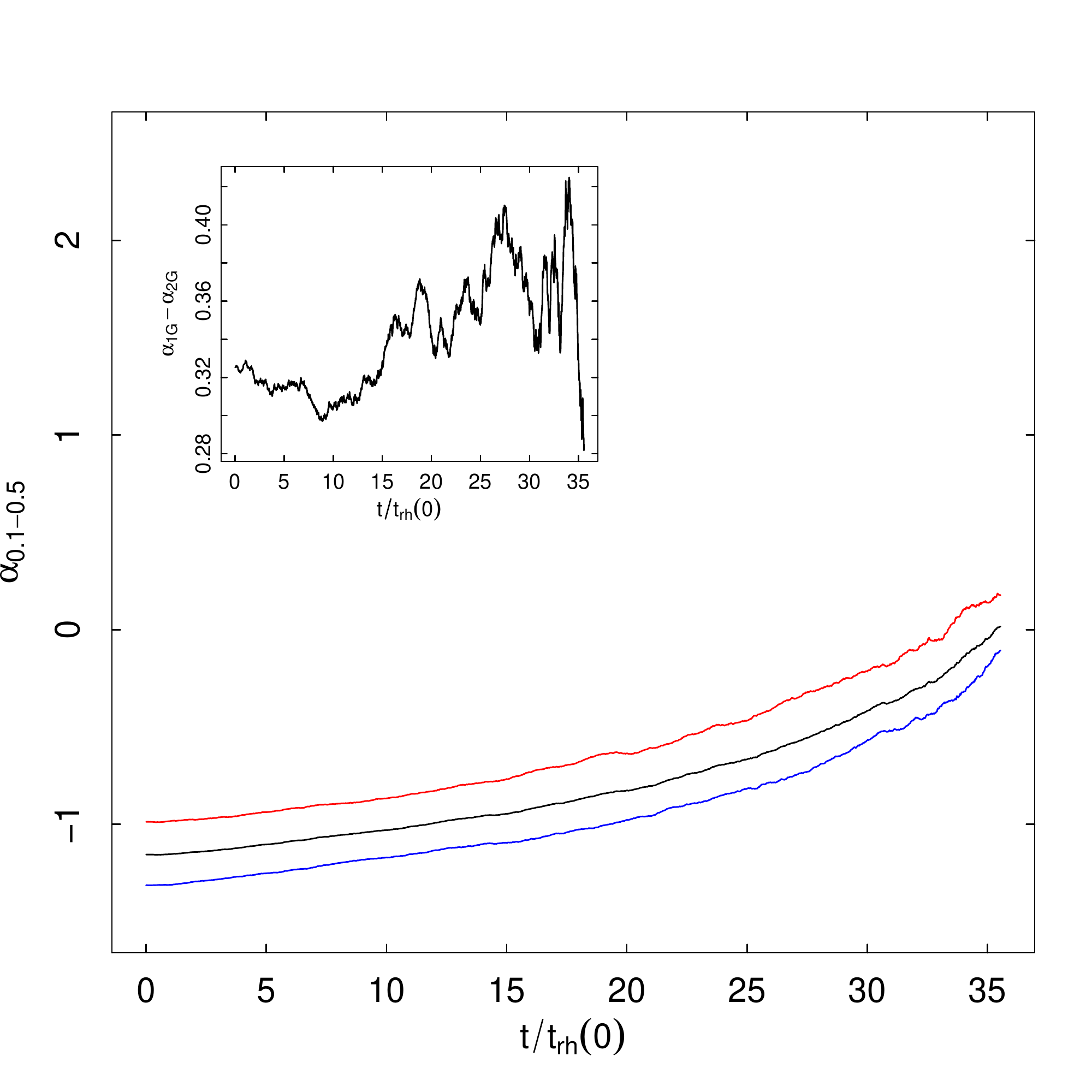}
  \includegraphics[width=5cm]{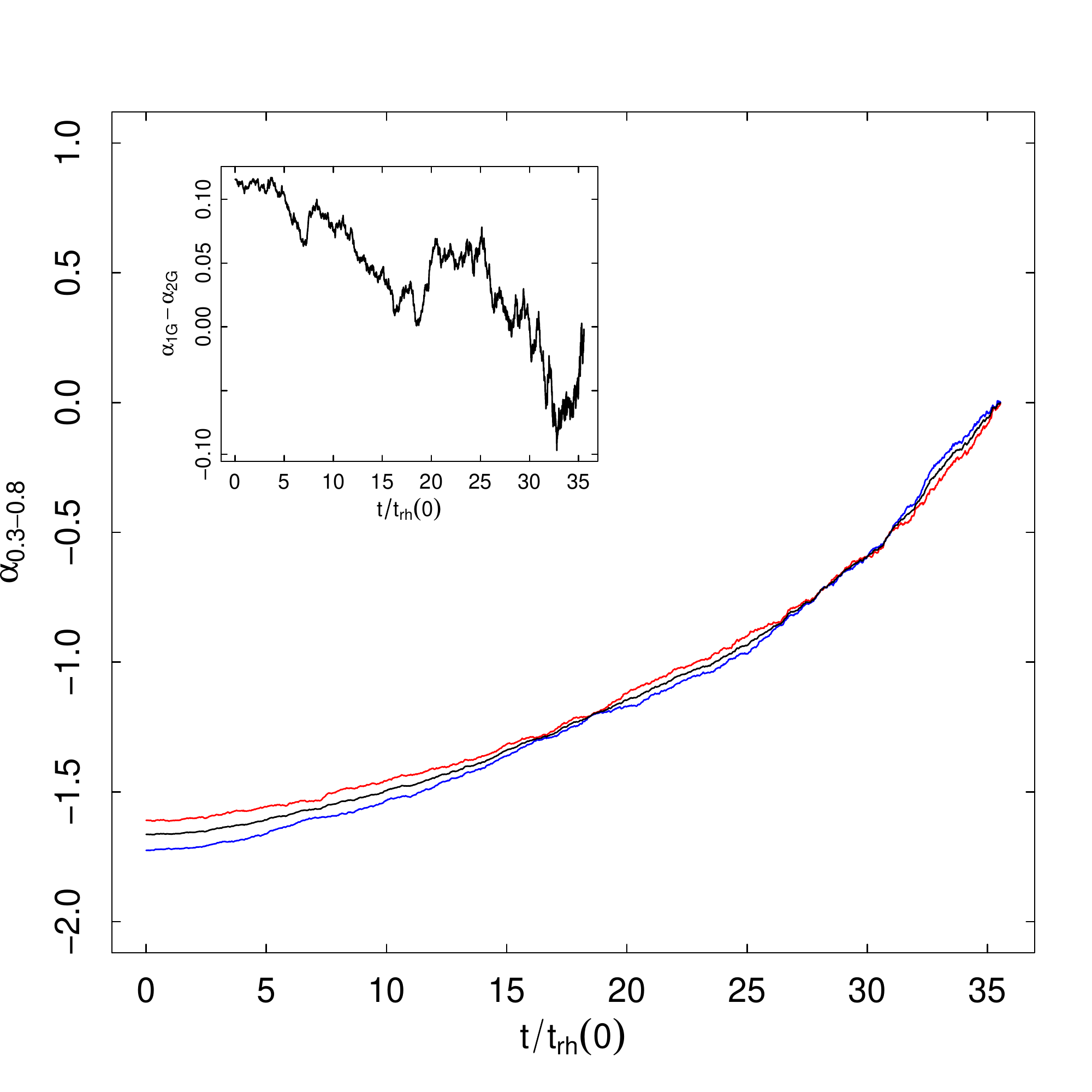}
}\\
\centering{
  \includegraphics[width=5cm]{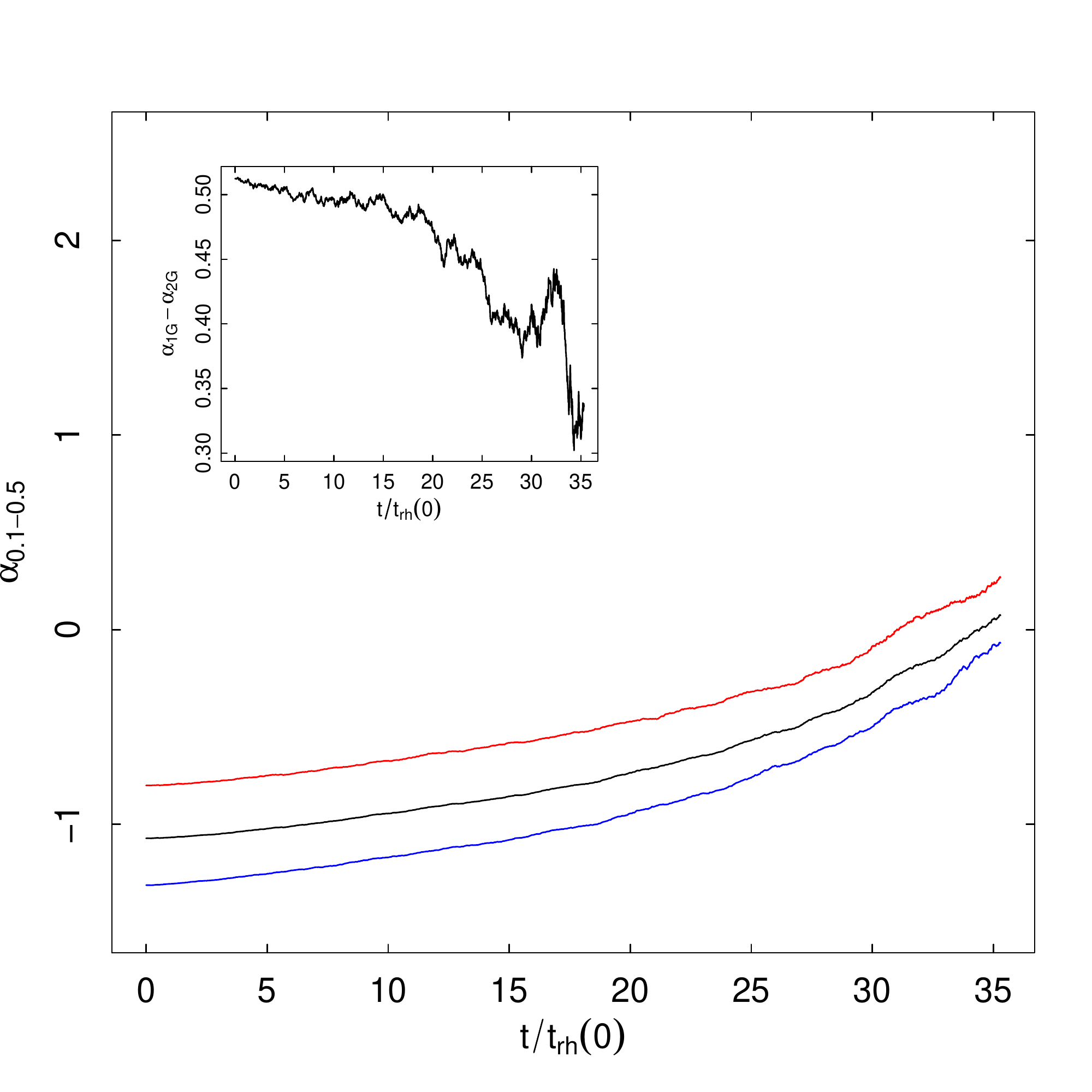}
  \includegraphics[width=5cm]{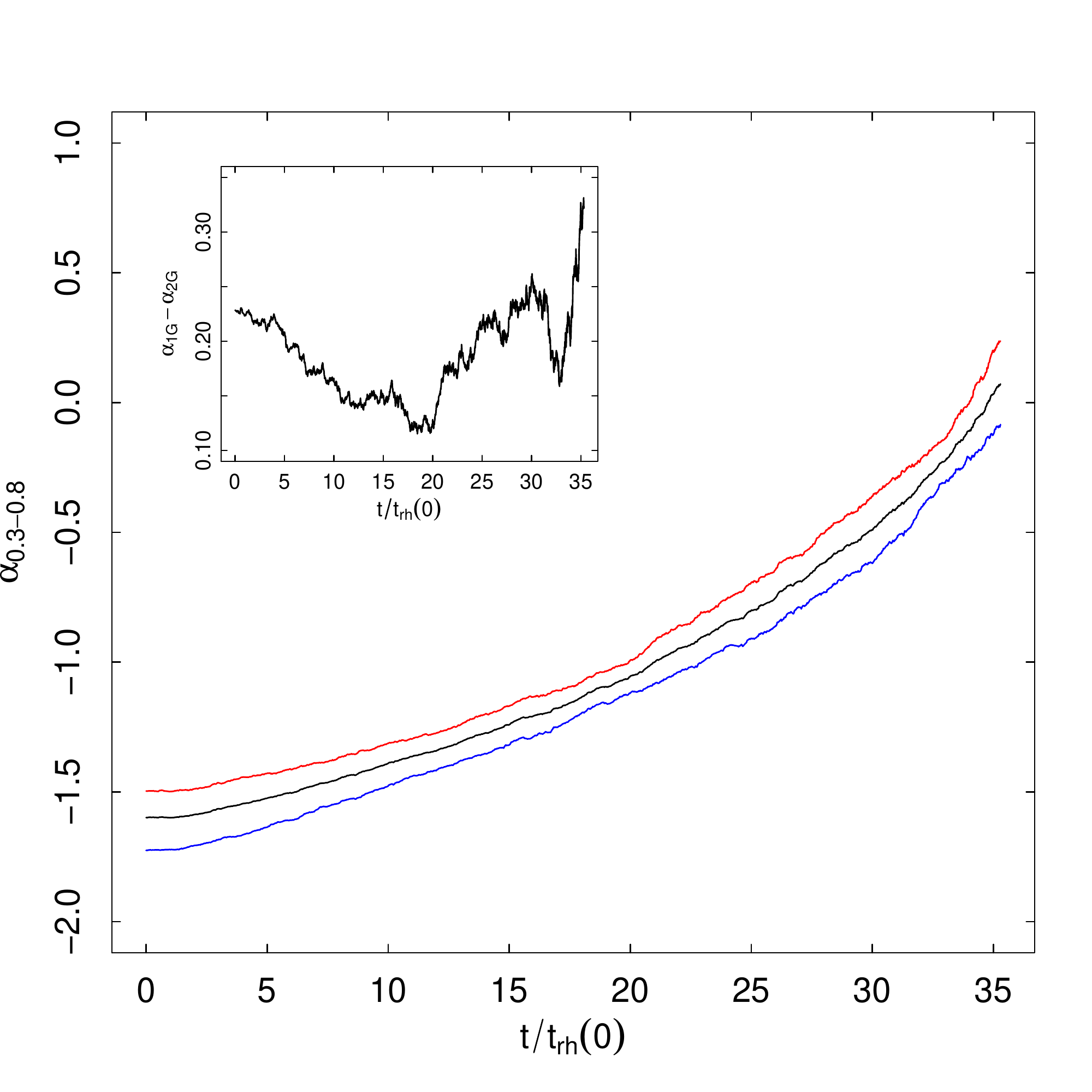}
}\\
\centering{
  \includegraphics[width=5cm]{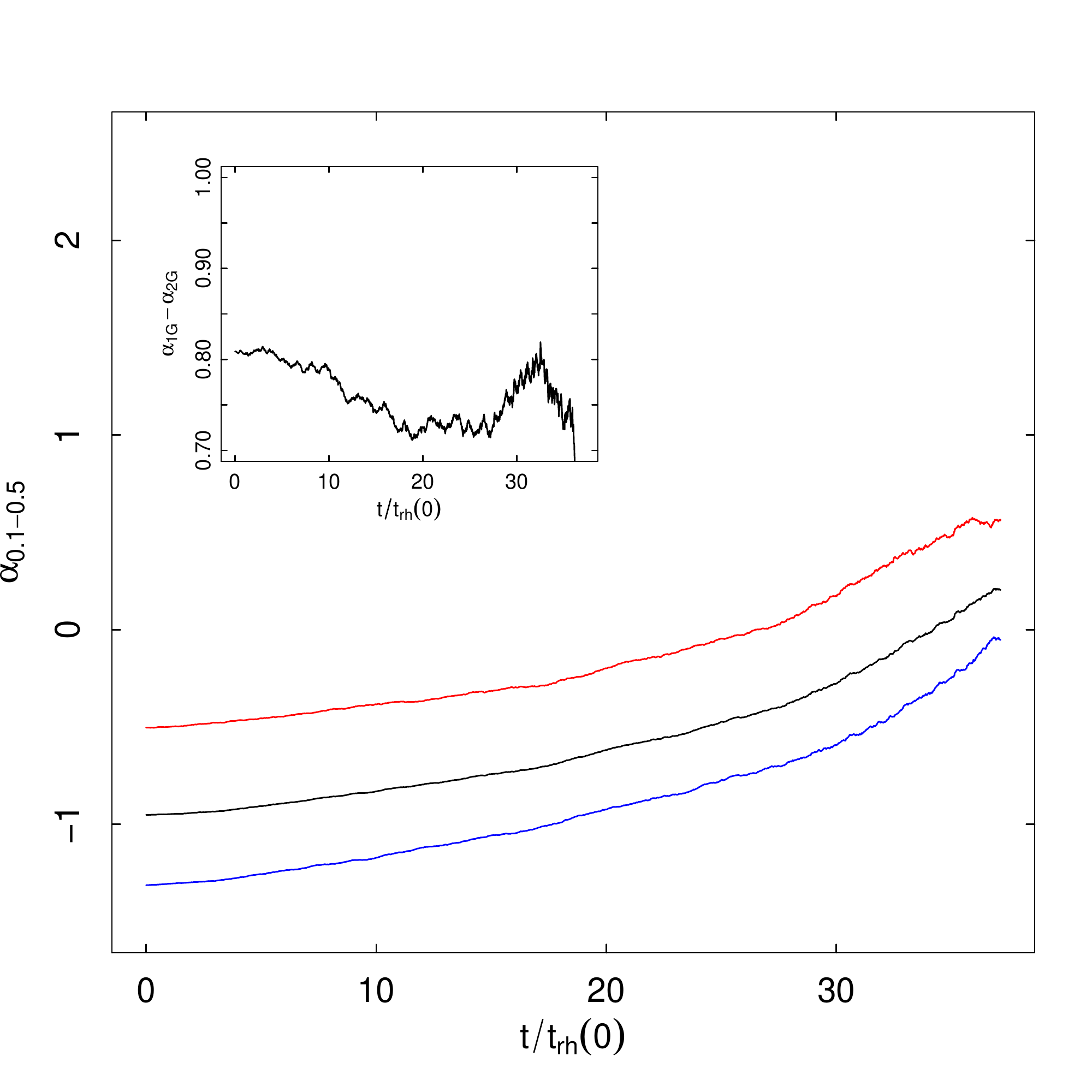}
  \includegraphics[width=5cm]{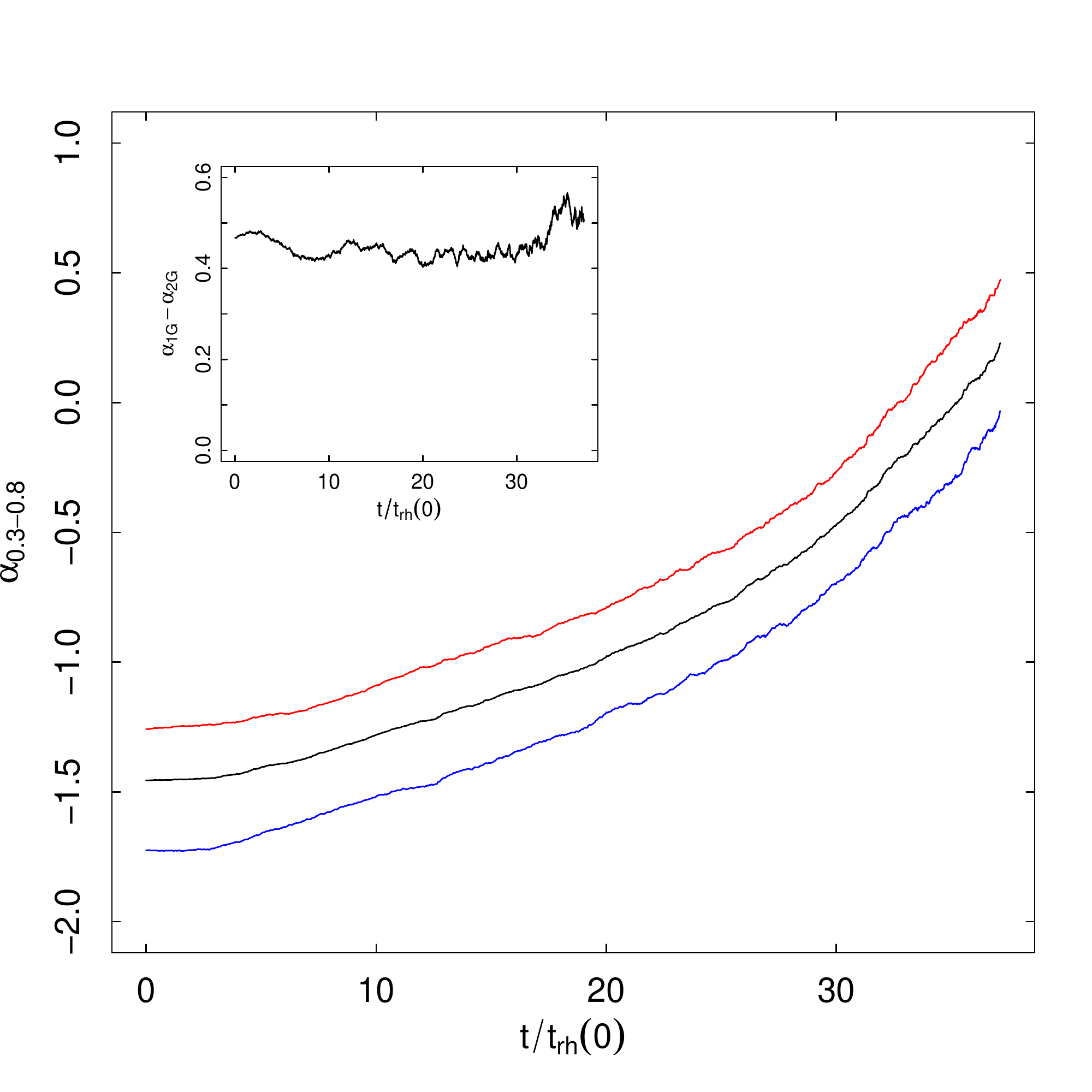}
}\\
\centering{
  \includegraphics[width=5cm]{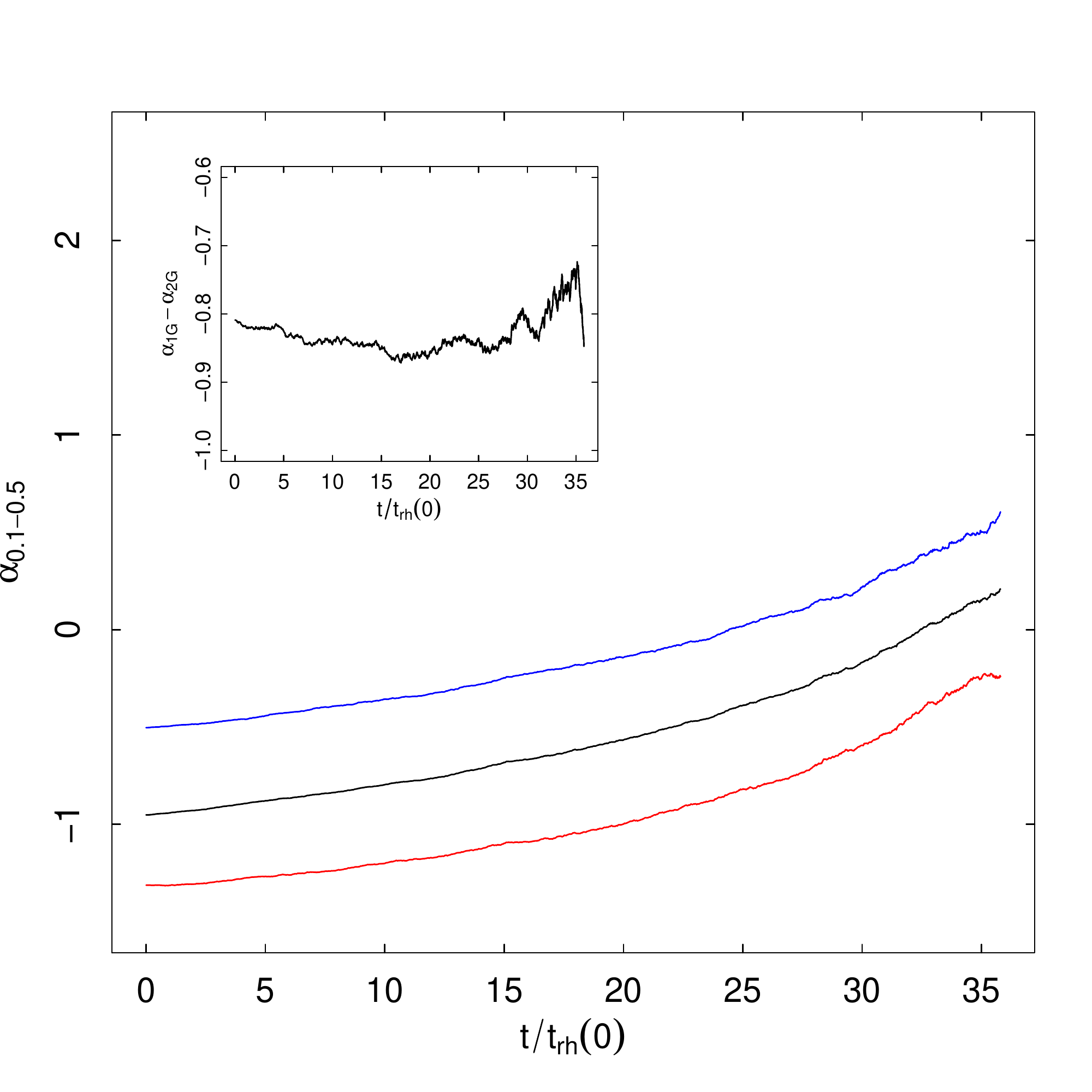}
  \includegraphics[width=5cm]{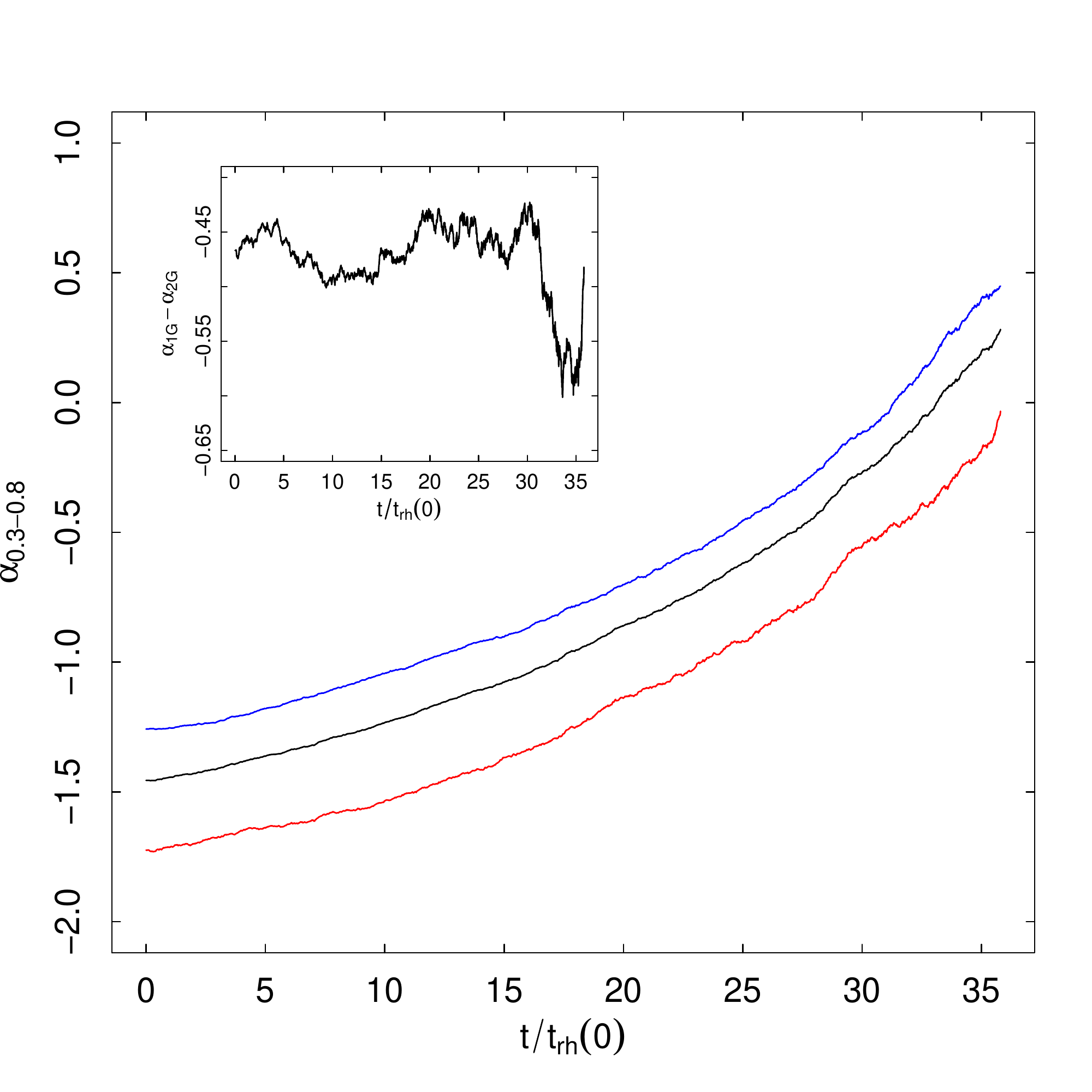}
}
  \caption{Time evolution of the slope of the global mass function for  2G stars (blue line), 1G stars (red line) and all stars (black line) for, from top to bottom panels, the K01R5a1 (first row), K01R5a08 (second row), and K01R5a05 (third row), and K01R5-2ga05 (fourth row) models. The panels on the left-hand column show the evolution of the slope of the mass function for stars with masses between $0.1 \msun$ and $0.5 \msun$; the panels on the right-hand colums show the evolution of the slope of the mass function for stars with masses between $0.3 \msun$ and $0.8 \msun$. 
The insets show the time evolution of the difference between the slope of the mass function of 2G stars and that of 1G stars for the mass range corresponding to each panel. Time is expressed in units of the initial half-mass relaxation time of the entire cluster. 
}
  \label{fig7}
\end{figure*}

Figure \ref{fig7} shows the time evolution of the global \alm$~$ and \aim$~$ of the 1G and the 2G populations for the four cases studied here: both the MF of the 1G and the 2G populations flatten as a result of the preferential loss of low-mass stars but the systems retain some memory of the initial differences in the IMFs of the two populations during the entire cluster evolution. Combining the results shown in this figure with those of section 3.1, we conclude that should observations reveal different global 1G and 2G MFs, this  would have to be a relic of differences in the IMF rather than the consequence of the effects of dynamical evolution.

\begin{figure*}
\centering{
  \includegraphics[width=7cm]{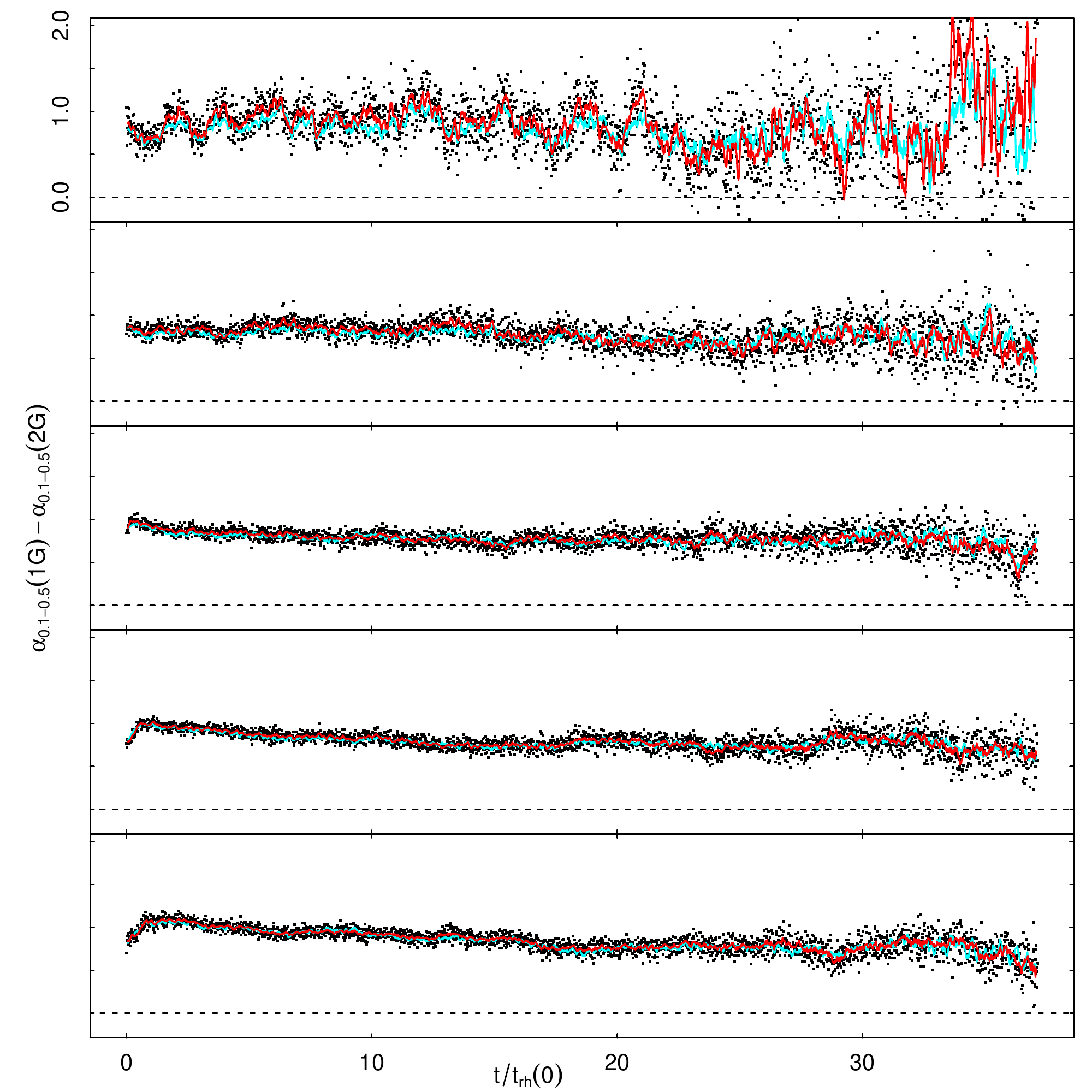}
  \includegraphics[width=7cm]{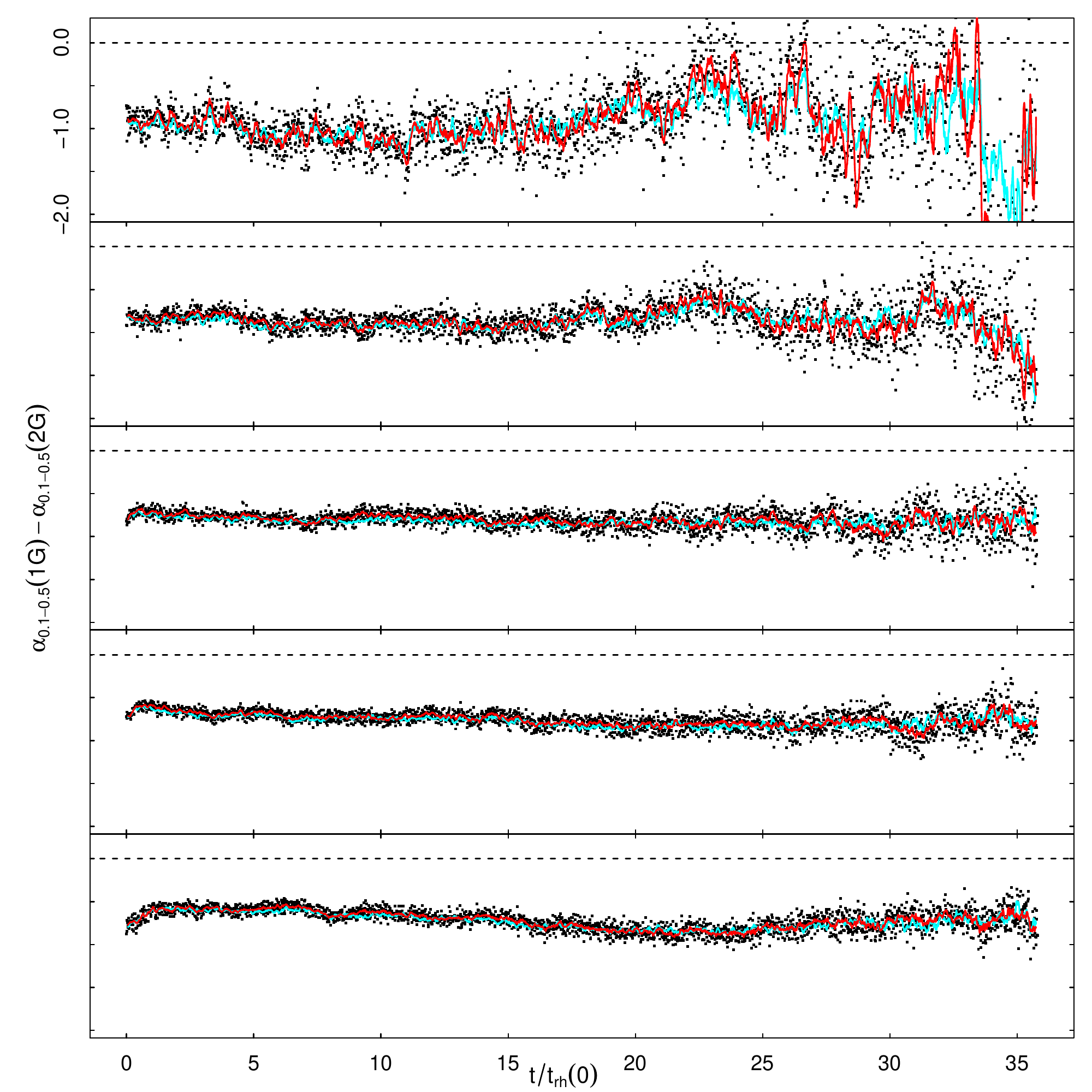}
}\\
\centering{
  \includegraphics[width=7cm]{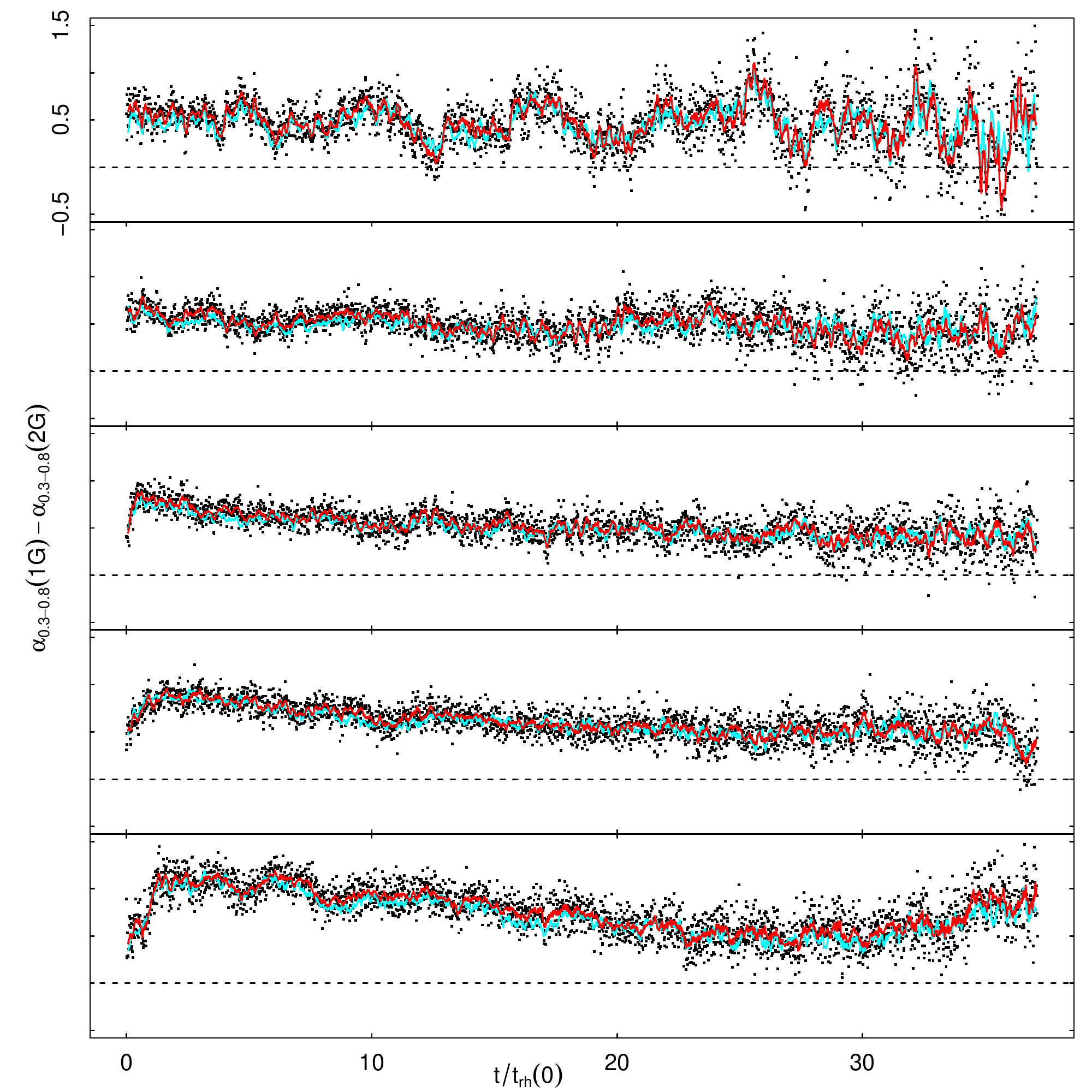}
  \includegraphics[width=7cm]{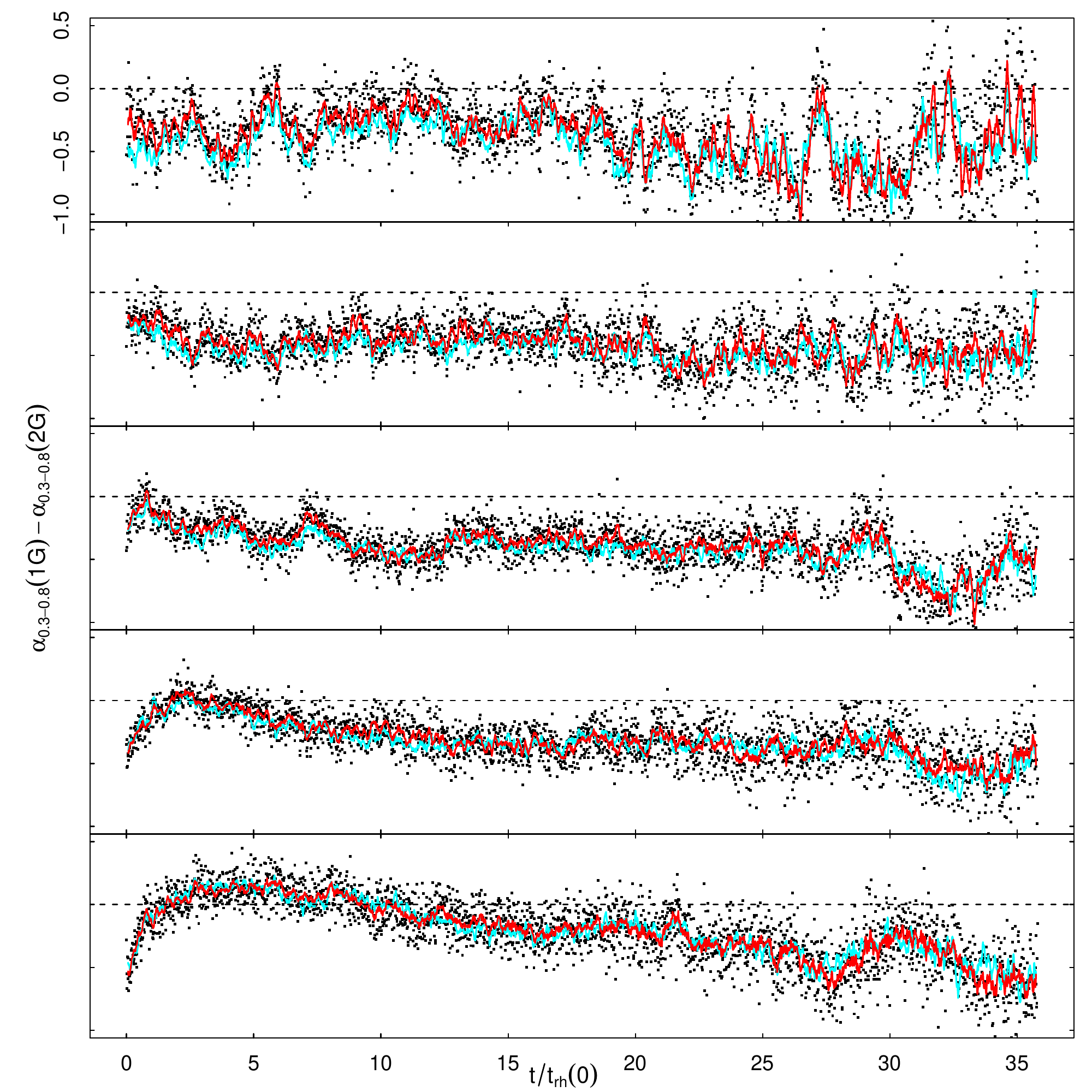}
}\\
\centering{
  \includegraphics[width=7cm]{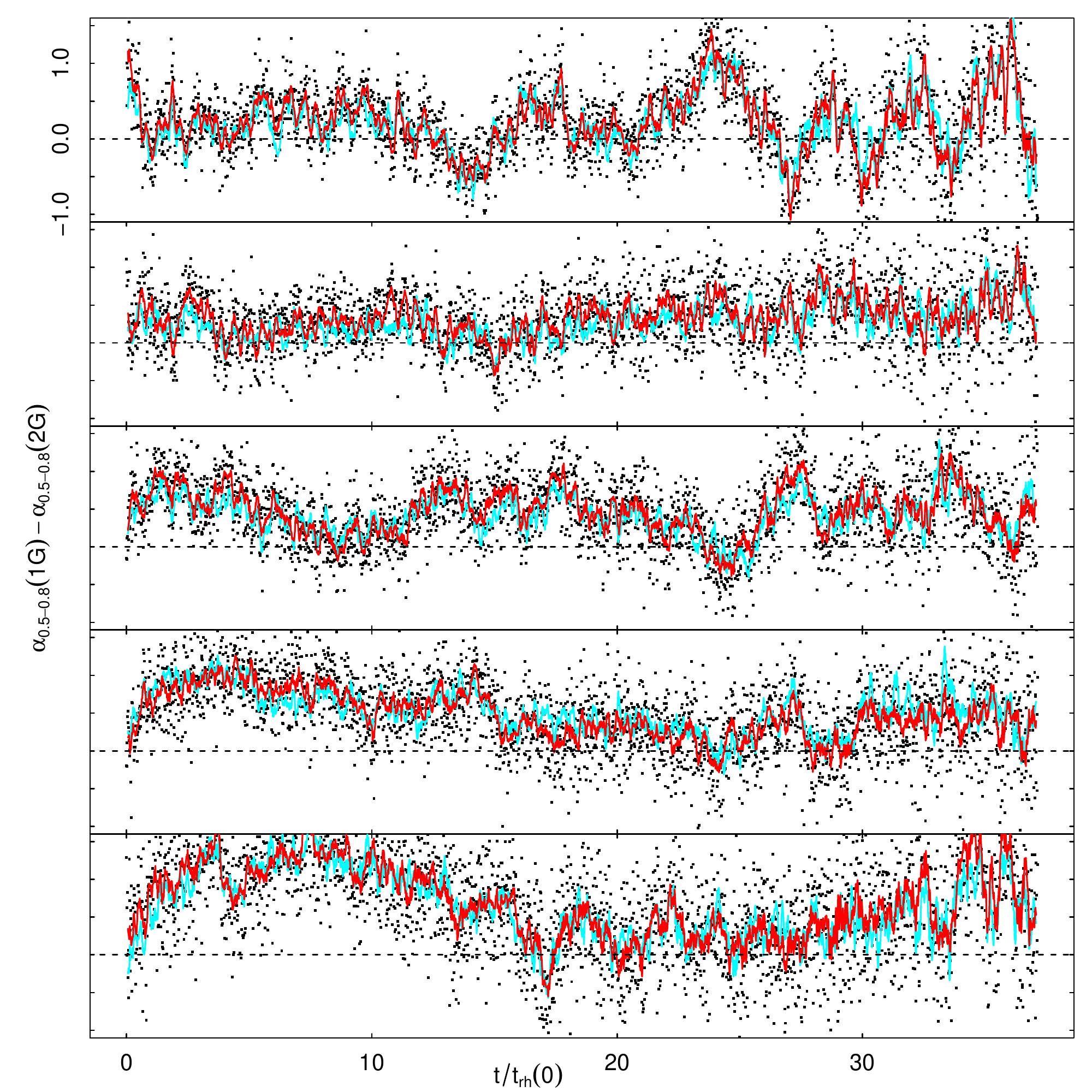}
  \includegraphics[width=7cm]{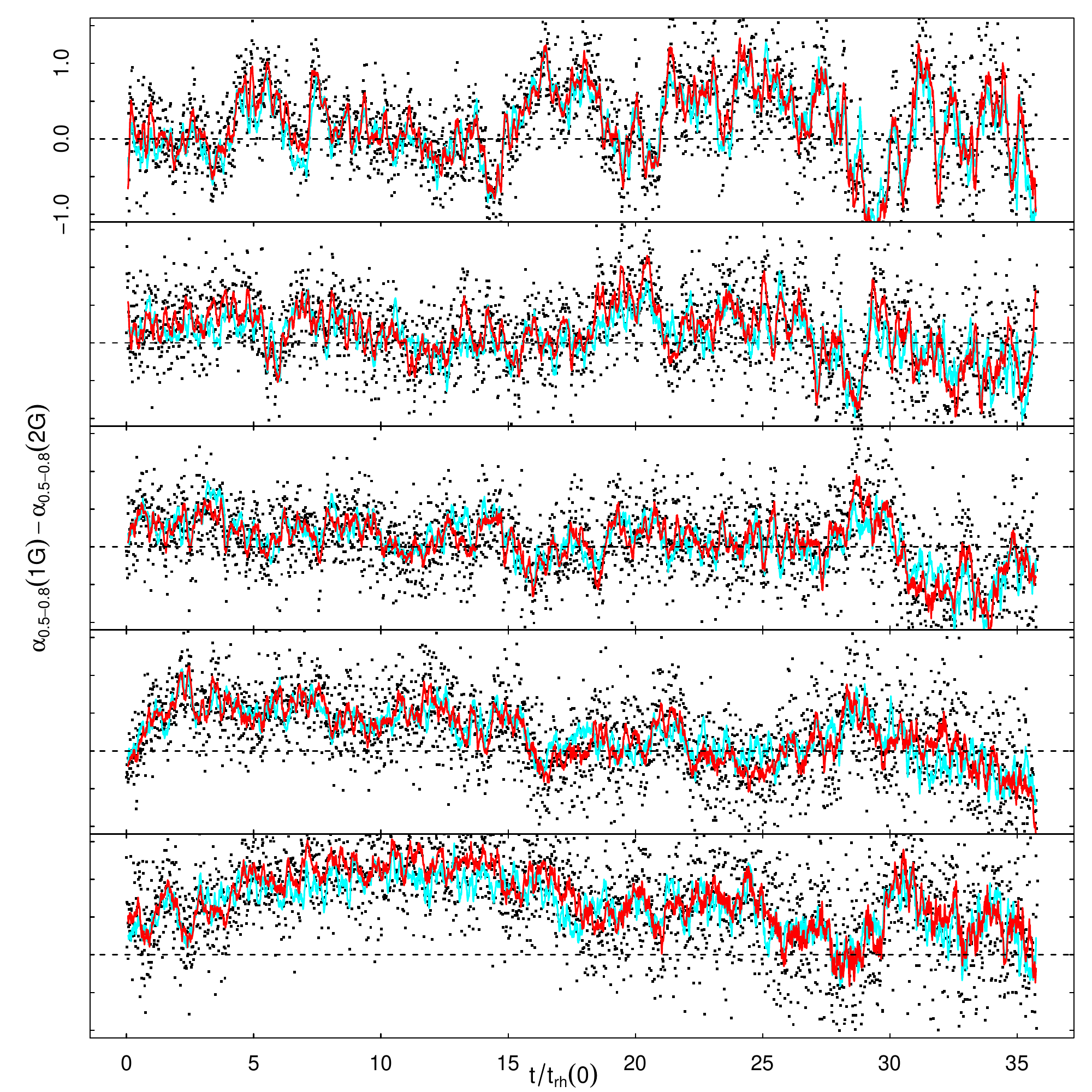}
}\\
  \caption{Same as Fig. \ref{fig6} for the model K01R5a05 (left-hand panels) and K01R5-2ga05 (right-hand panels).
}
  \label{fig8}
\end{figure*}

Fig.\ref{fig8} shows the time evolution of the local values of \alm(1G)-\alm(2G), \aim(1G)-\aim(2G), and \aum(1G)-\aum(2G)$~$  
(for the K01R5a05 and K01R5-2ga05 models): there are significant differences in the local MF of 1G and 2G stars at all distances from the cluster centre and for the entire cluster evolution. For these models differences between the 1G and 2G MF are present both in the local and in the global MF but, as already pointed out in section 3.2, we emphasize again that  caution is needed in the interpretation of differences in the local  MF as these can be present also for systems in which the 1G and the 2G populations have the same global MF. We point out, however, that in dynamically old clusters significant differences between the local 1G and 2G MFs are present only in those systems in which the 1G and 2G populations do not form with the same global IMF. We will further discuss this point in section 3.5.

\subsection{Spatial mixing of 1G and 2G stars}
\label{sectmix}
The differences in the evolution of the slope of the local MF measured at various distances from the cluster centre are a consequence of the effects of mass segregation in the 1G and  2G subsystems which have different initial structural properties and relaxation timescales. In the initially more compact 2G subsytem, mass segregation proceeds more rapidly than the more extended 1G system. Another imprint of these differences in the structure and relaxation timescales of the two subsystems  is found in the spatial mixing process.

\begin{figure}
\centering{
  \includegraphics[width=8cm]{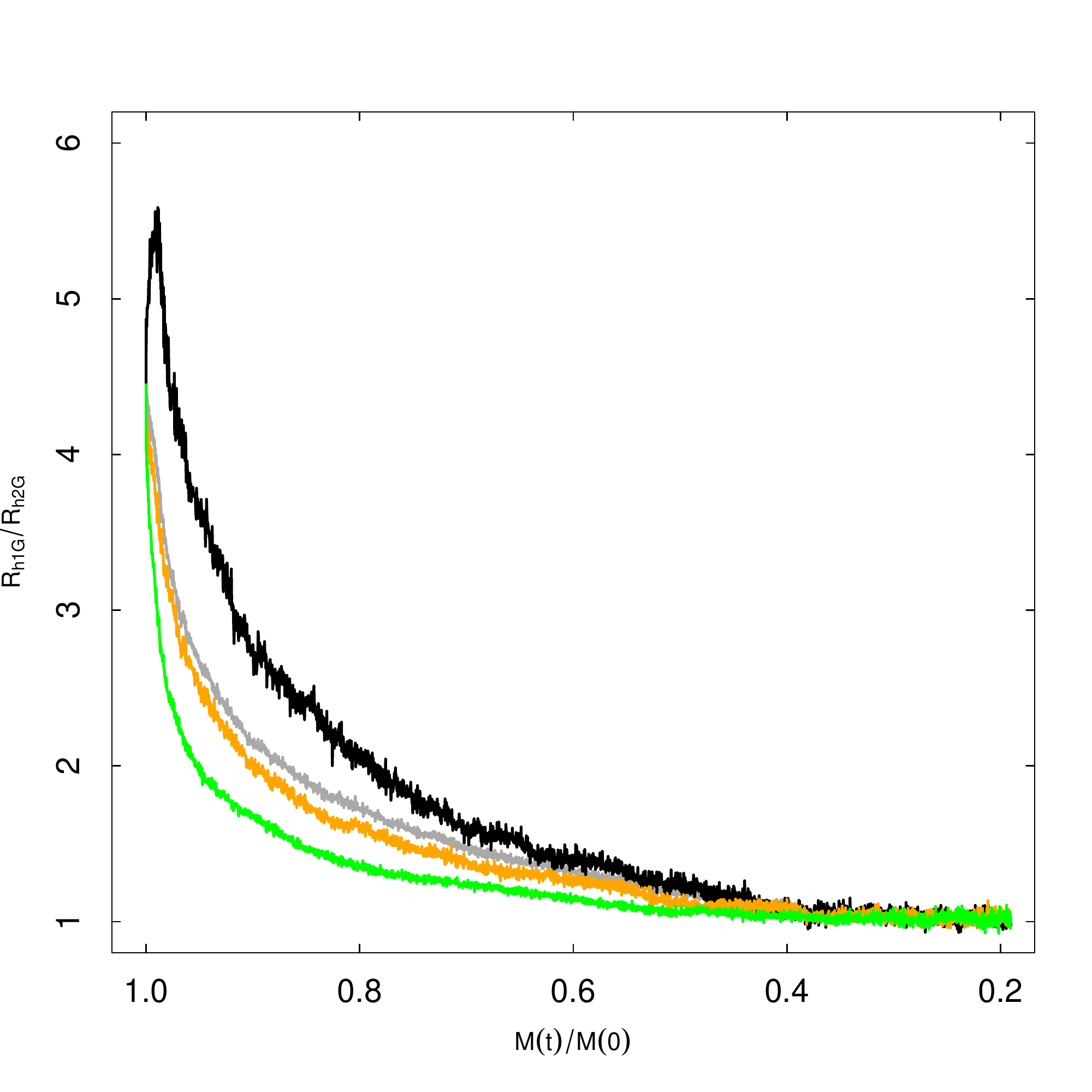}
}
  \caption{Evolution of the ratio of the 1G to the 2G 3D half-mass radius, $\rhratio$, as a function of the initial mass remaining in the cluster for the K01R5 model. The grey line shows the evolution of $\rhratio$ calculated using all the stars in the system. The green line shows $\rhratio$ for  stars with masses between 0.1 $\msun$ and 0.3 $\msun$, the orange line shows $\rhratio$ for  stars with masses between 0.3 $\msun$ and 0.6 $\msun$, and the black line shows $\rhratio$ for stars with masses between 0.6 $\msun$ and 0.85 $\msun$.
}
  \label{fig9}
\end{figure}

In Fig. \ref{fig9} we show the evolution of the ratio of the 1G to the 2G 3D half-mass radius, $\rhratio$  as a function of the fraction of the initial mass remaining in the cluster for the K01R5 system. This figure shows the  evolution of $\rhratio$ calculated using all the stars in the systems along with the evolution of $\rhratio$ for stars in three different mass ranges. The  evolution of $\rhratio$ for the entire cluster shows that, in agreement with the results of Vesperini et al. 2013 (see also Miholics et al. 2015), the two populations are mixed by the time the system has lost about 60 per cent of its initial mass (as already pointed out in Vesperini et al. 2013, we reiterate that here we refer to the mass loss due to two-body relaxation; any additional mass loss occuring early in the cluster evolution is not included here).

A new interesting aspect of the mixing process in multi-mass systems is however illustrated by the time evolution of $\rhratio$ for stars with different masses. Fig. \ref{fig9} clearly shows that the spatial mixing rate depends on the stellar mass: low-mass stars mix more rapidly than more massive stars and for the most massive bin considered $\rhratio$ even undergoes an initial increase. Although the dependence of the mixing rate on the stellar mass is not strong (and probably difficult to detect observationally), this figure shows an interesting dynamical manifestation of the multiscale structure of multiple-population clusters and the more rapid manifestation of the effects of segregation in the 2G subsystem: 2G massive stars rapidly sink toward the inner regions of the cluster leading to an initial increase in the difference between the 2G and 1G massive stars spatial distribution; on the other hand, 2G low-mass stars mix more rapidly as they migrate out of the inner regions towards the cluster outer regions. 
\subsection{Spatial mixing and the evolution of the mass function}
We conclude our analysis with a few remarks concerning the relationship between the evolution of the local and global MF of 1G and 2G stars and a cluster's dynamical evolution towards spatial mixing of 1G and 2G populations.
 
\begin{figure}
\centering{
  \includegraphics[width=8cm]{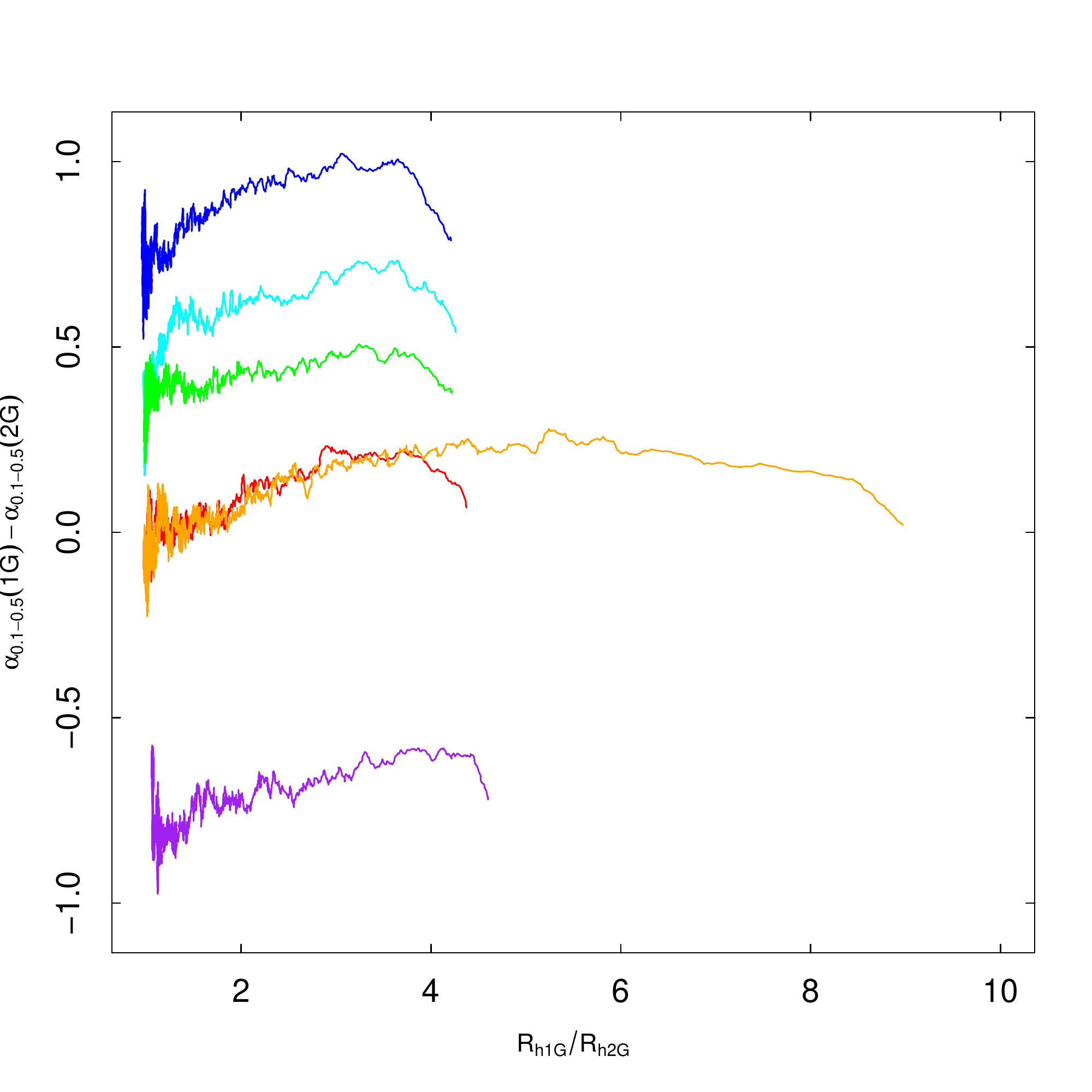}
  \includegraphics[width=8cm]{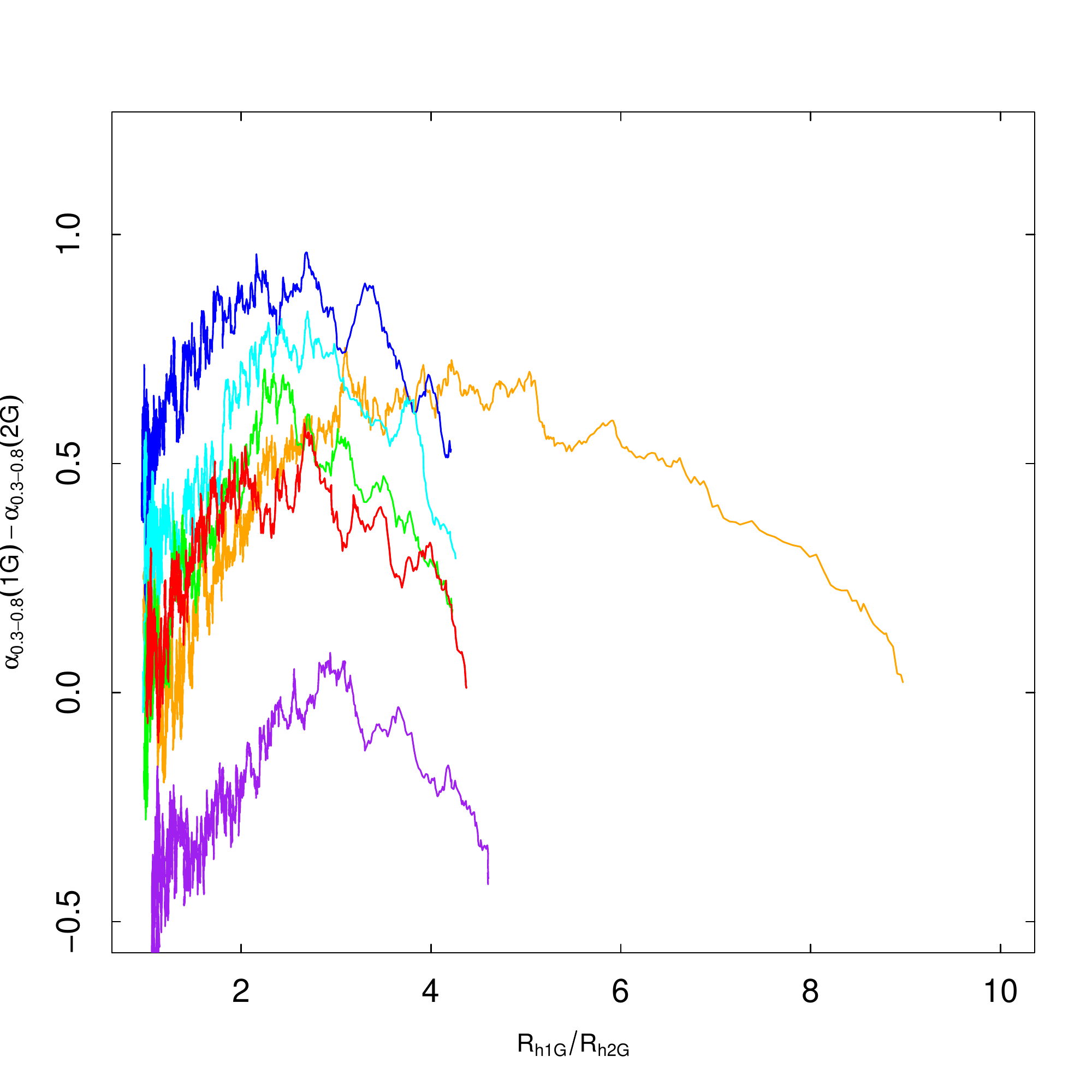}
}
  \caption{Evolution of the difference between the 1G and 2G slope of the MF in the mass range (0.1-0.5)$\msun$ (top panel) and (0.3-0.8)$\msun$ (bottom panel)
measured in a 3D shell between the 45 per cent and 65 per cent lagrangian radius as a function of the  ratio of the 1G to the 2G half-mass radius, $\rhratio$ for the K01R5 model (red line), K01R10 (orange line), K01R5a05 (blue line), K01R5a08 (cyan line), K01R5a1 (green line), K01R5-2ga05 (purple line).The values plotted are the rolling means (calculated using 10 points) of $\rhratio$ and of the difference between the slope of the 1G and 2G MF.
}
  \label{fig10}
\end{figure}

Fig.\ref{fig10} shows the evolution of the difference between the 2G and the 1G local values of \alm$~$(top panel) and \aim$~$(bottom panel)  measured between the (three-dimensional) 45 and the 65 per cent lagrangian radius versus $\rhratio$ for all the models studied in this paper. This figure illustrates the link between the different manifestations of  mass segregation and spatial mixing. As the cluster evolves towards complete spatial mixing, the difference in the local slope of the 1G and 2G MF initially increases and then decreases again late in the cluster evolution when the two populations are completely mixed. We point out  that during part of a cluster evolution the local slopes of the MFs of the 1G and the 2G population may differ from each other both in systems in which the two populations are characterized by the same global mass MF (K01R5 and K01R10) and for systems in which the two populations have different IMFs (K01R5a1, K01R5a08, K01R5a05, and K01R5-2ga05). 

An interesting point to emphasize here concerns the guidance provided by Fig.\ref{fig10} in how differences in the global MF of 1G and 2G stars might be identified.
As shown in this figure, for dynamically old systems that have already reached (or are close to) complete mixing, differences in the local slopes of the 1G and 2G MFs are present and non-negligible only in systems that started with different 1G and 2G global IMFs (although care must be used in considering possible differences due to noise). Observations of differences in the local MF in dynamically old and spatially mixed systems may therefore represent a powerful tool to reveal differences in the 1G and 2G global IMFs and PDMFs. We point out that, as clearly shown in Fig.\ref{fig10}, the extent of these differences in the local MFs depends on the mass range explored and its relationship with the mass range for which the PDMFs and the IMFs of the two populations differ from each other. 
This figure also shows that, irrespective of the cluster's dynamical age, large differences between the 1G and the 2G local MFs are more likely to be found in systems in which the two populations did not have the same IMF.

\begin{figure}
\centering{
  \includegraphics[width=8cm]{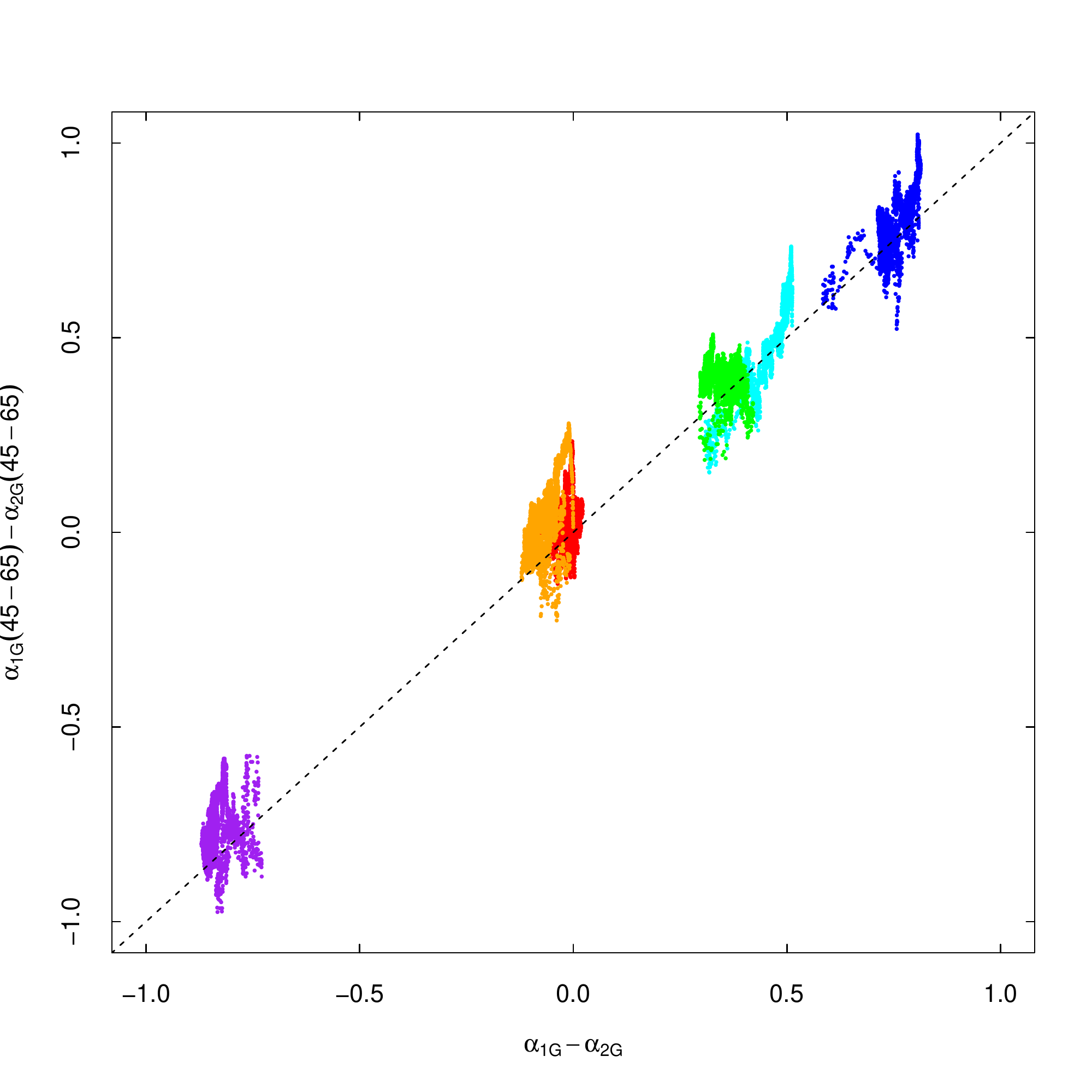}
  \includegraphics[width=8cm]{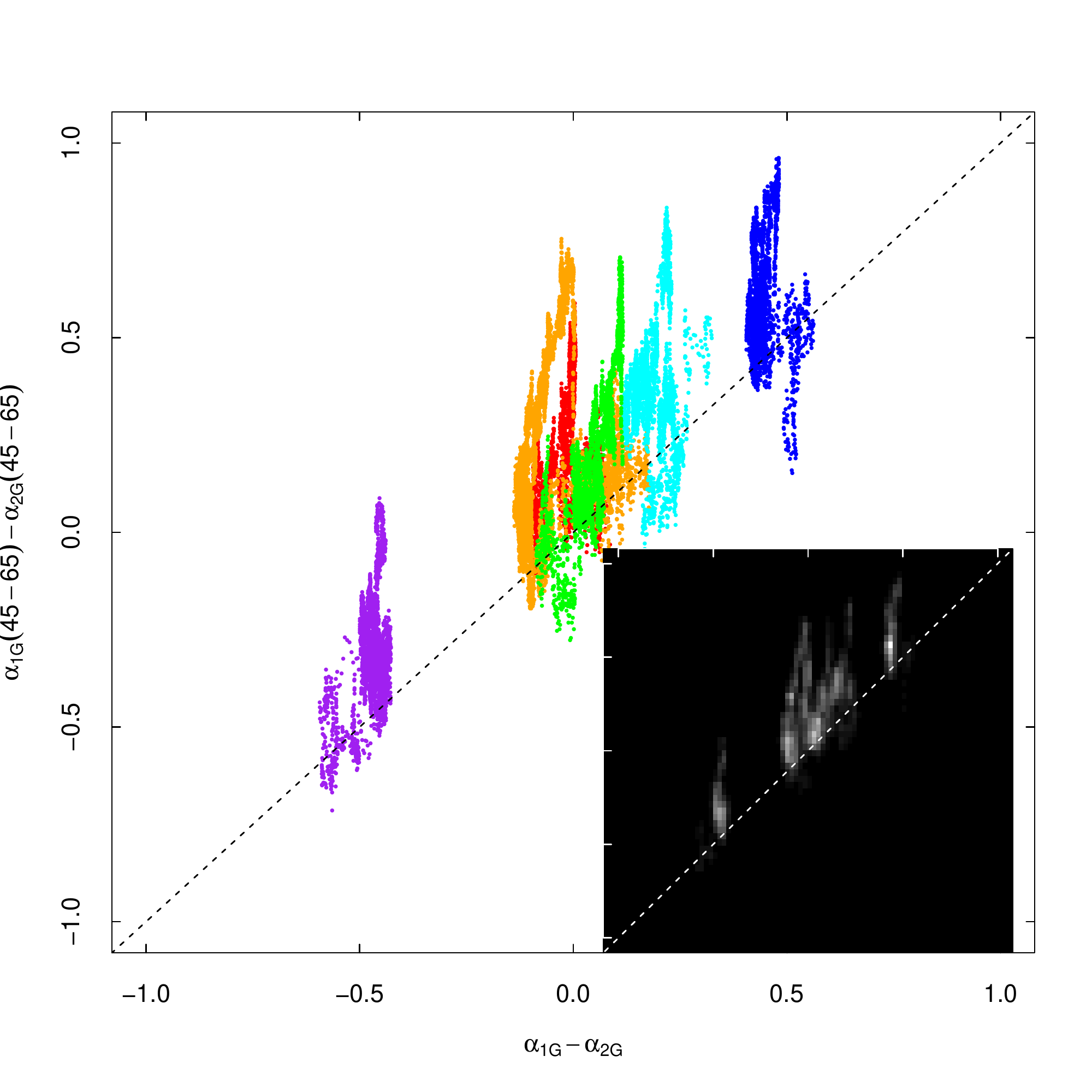}
}
  \caption{Difference between the 1G and 2G slope of the MF in the mass range (0.1-0.5)$\msun$ (top panel) and (0.3-0.8)$\msun$ (bottom panel)
measured in a 3D shell between the 45 per cent and 65 per cent lagrangian radius as a function of the  difference between the 1G and 2G slope of the global MF in the same mass range for the K01R5 model (red points), K01R10 (orange points), K01R5a05 (blue points), K01R5a08(cyan points), K01R5a1(green points), K01R5-2ga05 (purple points).The values plotted are the rolling means (calculated using 10 points). The inset in the bottom panel shows the density map of the same data shown in the main panel.
}
  \label{fig11}
\end{figure}

These points are further illustrated in Fig. \ref{fig11}. In the two panels of this figure we plot the differences between the slope of the 1G and 2G MF measured in a 3D shell between the 45 and the 65 per cent lagrangian radii versus the difference of the global slopes of the 1G and 2G MFs for stars with masses in the range $(0.1-0.5)\msun$ (top panel) and $(0.3-0.8)\msun$.
Although as discussed above and shown in Fig.\ref{fig10}, dynamically old and spatially mixed clusters are those for which differences in the local global slope of the 1G and 2G MF are more similar to each other, the top panel of  Fig.\ref{fig11} shows that when we focus on the specific stellar mass range (0.1-0.5)$\msun$ for which the 1G and 2G IMFs differed, the difference between the local slopes of the 1G and 2G MFs always provides a good indication of the differences in the global IMFs and PDMFs (we emphasize that the individual values of the MF slopes undergo a significant evolution as a cluster loses mass but the {\it difference} of the slopes does not vary significantly; see Fig.\ref{fig7}).
The bottom panel of Fig.\ref{fig11} shows that in the mass range (0.3-0.8)$\msun$ the difference between the local slopes of the 1G and 2G MFs undergoes a larger variation during a cluster's evolution than the difference between the global MF slopes. In this case dynamically old and spatially mixed clusters are those for which  differences in the local 1G and 2G MF slopes more closely resemble the differences in the global slopes of the 1G and 2G MFs (the density map in the inset is included to provide an indication of which regions of this plane are more likely to be populated during the evolution of the models presented in this paper).

We finally emphasize that, of course, dynamically young clusters that have not suffered any significant loss of stars and for which internal mass segregation has not altered the local MF slope also provide ideal targets for observations aimed at shedding light on the 1G and 2G global IMF from observations of the local PDMF.

$~$\\
\section{Conclusions}
The stellar MF is one of the fundamental properties characterizing a stellar population. A number of previous observational and theoretical investigations have explored the stellar MF of globular clusters, but the discovery of multiple stellar populations has raised many new challenges including a number of key questions concerning the MF, its possible dependence on the formation history of 1G and 2G stars as well as possible variations in the extent of the dynamical effects on the 1G and 2G MFs.
In this paper we have presented the results of a suite of N-body simulations aimed at studying the effects of dynamical evolution on the stellar MF in multiple-population globular clusters.

Different formation models of multiple stellar populations all agree that second-generation (2G) stars should form in the innermost regions of a more spatially extended first-generation (1G) cluster. We have explored the implications of the differences in the initial structural properties of the 1G and the 2G populations for the evolution of their stellar mass function (MF). Our simulations show that if 1G and 2G stars form with the same IMF, the {\it global} MFs of the two populations are affected similarly by dynamical evolution and mass loss: both the 1G and the 2G global MFs flatten as a result of the preferential escape of low-mass stars and no significant  differences between the 1G and the 2G MFs arise during the cluster's evolution (see Figs.\ref{fig1}-\ref{fig2}).

The differences in the initial structure of the 1G and 2G populations imply that until complete spatial mixing the two subsystems are characterized by different relaxation timescales and, therefore, that mass segregation proceeds at different rates for the two subsystems. As a consequence of the multiscale nature of the multiple-population cluster, even if the global MFs of the two populations are similar,  differences in the {\it local}  (measured in shells at  given radial distances from the cluster centre) MFs may arise during the cluster evolution (see Fig. \ref{fig6}). Differences in the local MFs, therefore, are not necessarily a manifestation of an actual difference in the global MFs of the two populations, but may rather reveal one of the consequences of the internal dynamical evolution of two subsystems with different initial structural properties.

Our simulations show that another consequence of the multiscale nature of multiple-population clusters is that 1G-2G spatial mixing rate depends on the stellar masses (see Fig.\ref{fig9}): low-mass stars mix more rapidly than more massive stars. We confirm the results of our previous study concerning the link between the degree of spatial mixing and the amount of mass lost due to the effects of internal relaxation (Figs.\ref{fig3}-\ref{fig4}). 

If the 1G and the 2G populations do not form with the same IMF, the dynamical effects on the evolution of the MF do not erase the initial MF differences during most of the cluster's evolution (see Fig.\ref{fig7}). Should observations reveal a difference between the global 1G and the 2G MFs this would have to be a relic of the formation/early dynamical history of the cluster rather than an effect of the cluster's long-term dynamical evolution. The differences between the local 1G and 2G MFs are in this case due both to dynamical effects and the actual differences in the global MFs of the two populations. Such differences in the local MFs are in general non-negligible also late in the evolution when the two populations are close to complete spatial mixing (Figs.\ref{fig10}-\ref{fig11}). Indeed observations of the local 1G and 2G MFs in dynamically old and spatially mixed clusters can reveal the presence of differences (or lack thereof) between the 1G and 2G global MFs. More in general, large differences between the local 1G and 2G MFs are likely to be associated with differences between the 1G and 2G global MFs for clusters with any dynamical age.
 In a future study we will further extend the analysis presented here to include primordial binaries; in particular we will explore the implications of the preferential disruption of 2G binaries (Vesperini et al. 2011, Hong et al. 2015, 2016) on the evolution of the MF taking into account the effects of unresolved binaries on the MF slope.

As the observational investigation of multiple stellar populations continues, it will be important to include in future observational studies of multiple stellar populations a significant investement in observations aimed at  studying  the MF of different populations. Future JWST observations may further extend  the identification of multiple populations down to the bottom of the main sequence, lead to a detailed characterization of multiple populations' MF, and provide key observational constraints to understand the possible role of the formation history and dynamical evolution in shaping the present-day MF.
\section*{Acknowledgments}

This research was supported in part by Lilly Endowment, Inc., through its support for the Indiana University Pervasive Technology Institute, and in part by the Indiana METACyt Initiative. The Indiana METACyt Initiative at IU is also supported in part by Lilly Endowment, Inc.
Support from grant HST-AR-13273.01-A is acknowledged. Support for Program number HST-AR-13273.01-A was provided by NASA through
a grant from the Space Telescope Science Institute, which is operated by the
Association of Universities for Research in Astronomy, Incorporated, under
NASA contract NAS5-26555.
JH acknowledges support from the China Postdoctoral Science Foundation, Grant No. 2017M610694.

\end{document}